\documentclass[usenatbib,usegraphicx,letterpaper]{mn2e}
\usepackage[totalwidth=500pt,totalheight=680pt]{geometry}

\usepackage{amssymb}
\usepackage{epsfig}
\usepackage{amsmath}
\usepackage{color}

\newcommand{\apjl}{ApJL~}
\newcommand{\apjs}{ApJS~}

\newcommand{\unit}[1]{\mathrm{#1}}

\newcommand{\wprp}{w_{\mathrm{p}}(r_{\mathrm{p}})}

\newcommand{\Mpc}{\unit{Mpc}} 
\newcommand{\kpc}{\unit{kpc}}
\newcommand{\Gyr}{\unit{Gyr}} 
\newcommand{\Myr}{\unit{Myr}}

\newcommand{\hmpc}{h^{-1}\mathrm{Mpc}}
\newcommand{\hmpcvol}{h^{-3}\mathrm{Mpc^{3}}}
\newcommand{\hkpc}{h^{-1}\mathrm{kpc}}
\newcommand{\hMsun}{h^{-1}M_{\odot}}

\newcommand{\Mvir}{M_\mathrm{vir}} 
\newcommand{\Mhost}{M_\mathrm{host}} 
\newcommand{\Mhalo}{M_\mathrm{halo}} 
\newcommand{\Mstar}{M_{\ast}}

\newcommand{\Rvir}{R_\mathrm{vir}}

\newcommand{\Msun}{M_{\odot}}

\newcommand{\vmax}{\mathrm{V}_\mathrm{max}}

\newcommand{\vpeak}{\mathrm{V}_\mathrm{peak}}

\newcommand{\vacc}{\mathrm{V}_\mathrm{acc}}

\newcommand{\zform}{z_{\mathrm{form}}}
\newcommand{\zacc}{z_{\mathrm{acc}}}
\newcommand{\zchar}{z_{\mathrm{char}}}
\newcommand{\zstarve}{z_{\mathrm{starve}}}
\newcommand{\xhalo}{X_{\mathrm{halo}}}
\newcommand{\xgal}{X_{\mathrm{gal}}}

\newcommand{\PsdssMstar}{P_{\mathrm{SDSS}} ( \mathrm{sSFR} | M_{\ast} )}

\newcommand{\SMcat}{{M}_{\ast}^{9.8}}

\newcommand{\beq}{\begin{equation}}
\newcommand{\eeq}{\end{equation}}
\newcommand{\beqray}{\begin{eqnarray}}
\newcommand{\eeqray}{\end{eqnarray}}

\newcommand{\ben}{\begin{enumerate}}
\newcommand{\een}{\end{enumerate}}
\newcommand{\bit}{\begin{itemize}}
\newcommand{\eit}{\end{itemize}}

\usepackage{epsfig}  \usepackage{graphicx}   \usepackage{rotating}

\begin{document}

\title[Predicting Galaxy Star Formation Rates]
{Predicting Galaxy Star Formation Rates via the Co-evolution of Galaxies and Halos}


\author[Watson et al.]
{Douglas~F.~Watson$^{1}$\thanks{NSF Astronomy \& Astrophysics Postdoctoral Fellow}\thanks{email: dfwatson@kicp.uchicago.edu}, Andrew~P.~Hearin$^{1,2,3}$, Andreas~A.~Berlind$^{4}$, Matthew~R.~Becker$^{5,6,7}$,
\newauthor
Peter~S.~Behroozi$^{8}$, Ramin~A.~Skibba$^{9}$,  Reinabelle Reyes$^{1}$, Andrew~R.~Zentner$^{10,11}$, 
\newauthor
Frank~C.~van den Bosch$^{12}$ \\
$^1$Kavli Institute for Cosmological Physics, 5640 South Ellis Avenue, The University of Chicago, Chicago, IL \\
$^2$Fermilab Center for Particle Astrophysics, Fermi National Accelerator Laboratory, Batavia, IL, \\
$^3$Yale Center for Astronomy \& Astrophysics, Yale University, New Haven, CT\\
$^4$ Department of Physics and Astronomy, Vanderbilt University, Nashville, TN\\
$^5$ SLAC National Accelerator Laboratory, Menlo Park, CA 94025 \\
$^6$ Kavli Institute for Particle Astrophysics and Cosmology, Stanford, CA 94309, USA \\
$^7$ Department of Physics, School of Humanities and Sciences, Stanford University, Stanford, CA 94309, USA \\
$^8$  Space Telescope Science Institute, 3700 San Martin Drive, Baltimore, MD 21218 \\
 $^9$ Department of Physics, Center for Astrophysics and Space Sciences, University of California, 9500 Gilman Dr., La Jolla, San Diego, CA 92093 \\
 $^{10}$ Department of Physics and Astronomy, University of Pittsburgh, Pittsburgh, PA 15260 \\
 $^{11}$ Pittsburgh Particle physics, Astrophysics and Cosmology Center (PITT PACC) \\
 $^{12}$ Department of Astronomy, Yale University, P.O. Box 208101, New Haven, CT \\
}

\maketitle

\begin{abstract}

In this paper, we test the {\em age matching} hypothesis that the star
formation rate (SFR) of a galaxy of fixed stellar mass is determined
by its dark matter halo formation history, and as such, that more
quiescent galaxies reside in older halos.  This simple model has been
remarkably successful at predicting color-based galaxy statistics at
low redshift as measured in the Sloan Digital Sky Survey (SDSS).  To
further test this method with observations, we present new SDSS
measurements of the galaxy two-point correlation function and
galaxy-galaxy lensing as a function of stellar mass and SFR, separated
into quenched and star-forming galaxy samples.  We find that our age
matching model is in excellent agreement with these  new measurements.
We also employ a galaxy group finder and show that our model is able
to predict: (1) the relative SFRs of central and satellite galaxies,
(2) the SFR-dependence of the radial distribution of satellite galaxy
populations within galaxy groups, rich groups, and clusters and their
surrounding larger scale environments, and (3) the interesting feature
that the satellite quenched fraction as a function of projected radial
distance from the central galaxy exhibits an $\sim \mathrm{r}^{-.15}$
slope, independent of environment. The accurate prediction for the
spatial distribution of satellites is intriguing given the fact that
we do not explicitly model  satellite-specific processes after infall,
and that in our model the virial radius does not mark a special
transition region in the evolution of a satellite, contrary to most
galaxy evolution models.  The success of the model suggests that
present-day galaxy SFR is strongly correlated with halo mass assembly
history.

\end{abstract}



\section{INTRODUCTION}
\label{sec:intro}

One of the principal goals of galaxy evolution theory is to understand
the connection between the properties of galaxies and their host dark
matter halos.  There is now a well-established relation between the
stellar mass of a galaxy and the mass of the halo in which it resides
\citep[e.g.,][]{yang12,leitner12,wang_etal12,moster13,behroozi13b,kravtsov_size_Rvir13,kravtsov_smhm_14}.
Moreover, the fact that the stellar mass-to-halo mass connection remains
tight across cosmic time
\citep{conroy_wechsler09,behroozi13,watson_conroy13,lu_etal14}  suggests that
there are likely further links between halo properties and the star
formation rate (SFR) of galaxies. With this as motivation, the aim of
the present work is to address the following question: is there a
simple link between  the SFR of galaxies and the dark side of the
universe?

The complex nature of star formation in galaxies indicates that  the
relationship between the SFR of a galaxy and the properties of its
host dark matter halo may be complicated.  First, at {\em fixed
  luminosity} or {\em fixed stellar mass}, there exists a clear
bimodality in the distribution of galaxy color/SFR, with distinct
red/quenched and blue/star-forming populations
\citep{blanton03,baldry04,bell04,blanton05,cooper06,wyder07,wetzel_etal11,cooper12}.
Additionally, galaxy color/SFR depends on environment: denser
environments, such as rich groups and clusters, are populated by
significantly more red sequence galaxies than actively star-forming
ones
\citep{balogh99,blanton05,weinmann06b,weinmann09,peng_etal10,peng_etal12,carollo_etal12,tal_etal14}.
Furthermore, the specific processes that attenuate SFR in a 'central'
galaxy (the galaxy at the minimum of the halo potential well) may be
distinct from those governing the 'satellite' galaxies orbiting the
central \citep{vdbosch_08}.  Finally, the SFR/color dependence of
galaxy location within the cosmic web also manifests in measurements
of two-point statistics; as a function luminosity or stellar mass,
red/quenched galaxies exhibit stronger clustering than blue/star
forming galaxies
\citep[e.g.,][]{norberg02,zehavi02,zehavi05a,li06,zehavi11,yang12,mostek12,guo_etal14}.
Such observed complexities may lead one to conclude that complicated
modeling of the physics governing the quenching of galaxies is
required to reproduce observed galaxy statistics.

However, in a pair of recent papers introducing the {\em age matching}
formalism \citep[][hereafter Papers I $\&$ II,
  respectively]{HW13a,HW13b},  it was shown that in fact a very simple
model for galaxy color can account for the rich variety of trends
exhibited by galaxies in the low-redshift universe.  The central
hypothesis of age matching is that at fixed luminosity (or fixed
stellar mass), galaxy color  is in monotonic correspondence with a
proxy for halo age, at fixed halo maximum circular velocity $\vmax$.  In Paper
I, this prescription was shown to accurately reproduce the
observed $g-r$ color-dependent clustering of galaxies in the Sloan
Digital Sky Survey \citep[SDSS:][]{york00a} for the
luminosity-selected galaxy samples presented in \citet{zehavi11}, as
well as the scaling between $g-r$ color and host halo  mass.  In Paper
II, similar success of the age matching formalism was demonstrated for
model predictions of new measurements of both SDSS clustering and
galaxy-galaxy lensing as a function of stellar mass and $g-r$ color.

However, $g-r$ color and star formation activity are not perfectly
correlated.  For instance, galaxies that are actively forming stars
can often appear red due to the ubiquitous presence of dust
\citep[e.g.,][]{stein_soifer83,maller_etal09,masters_etal10}.  Furthermore, $g-r$
color is the convolution of many physical properties of galaxies,
including: stellar age, metallicity, and instantaneous SFR.  Thus, a
model for the color-dependence of galaxy location within the cosmic
web may not smoothly translate to a model for the SFR dependence.  In
the present study, we demonstrate how age matching, without
modification to the technique introduced in Papers I $\&$ II, is
equally successful at reproducing new SDSS measurements of stellar
mass- and SFR-dependent clustering and galaxy-galaxy lensing.
As we will demonstrate in a forthcoming paper (Watson, Skibba $\&$ Hearin 2014, in prep) studying {\em marked} correlation functions
\citep{skibba06,skibbamarkedCF13}, this simultaneous success of our model 
is primarily due to the surprising observational fact that the two-point function 
is almost entirely insensitive to the choice of SFR or $g-r$ as a star formation indicator.

Additionally, in this paper we take a sharp focus on the population of
satellite galaxies.   While satellites are in the minority by
number, the physics governing satellite galaxy SFR is a key ingredient to painting a
complete picture of the theory of galaxy evolution.  Satellite
galaxies can be subject to a number of complex processes which are
believed to stifle their star formation as they orbit within the
gravitational potential well of their host halo. These include the
removal of cold gas from the disc due to ram pressure
\citep{gunn_gott72}, the stripping of the surrounding hot gas
reservoir, known as `strangulation' \citep{larson80}, disruption of
satellite galaxies due to tidal stripping \citep{purcell_etal07,
  watson_etal12b}, and `harassment' by gravitational interactions with
other nearby galaxies \citep{moore_etal98}. 

In age matching there is {\em no} explicit modeling of such
post-accretion processes.  And yet, we will demonstrate that this
remarkably simple model accurately predicts the radial profiles of
star-forming and quenched satellite galaxies within and around the
environment of groups, rich groups, and clusters.  As discussed in
\S~\ref{sec:discussion}, the success of age matching at reproducing
these trends indicates that in much of the literature on satellite
evolution, the influence of post-accretion processes on quenching
satellites has been over-estimated.

The paper is laid out as follows.  In \S~\ref{sec:data} we describe
the data, simulation and halo catalogs incorporated throughout this
work.  An overview of the age matching and the more generic ``conditional abundance matching'' formalism is given in
\S~\ref{sec:model}.  In \S~\ref{sec:results} we present our main
results.  Specifically, in \S~\ref{subsec:2PCF_gg_results} we show our
model predictions for new measurements of the SFR-dependent two-point
correlation function and galaxy-galaxy lensing signal.  In
\S~\ref{subsec:sat_results} we study the spatial properties of
star-forming and quenched satellite galaxies within and around halos.
In \S~\ref{sec:discussion} we provide a discussion and interpretation
of our findings.  We conclude in \S~\ref{sec:summary} with a brief
summary of our primary results.  Throughout this work we assume a flat
$\Lambda$CDM cosmological model with $\Omega_{\mathrm{m}}=0.27$ and
Hubble constant $H_0=70$ km s$^{-1}$ Mpc$^{-1}$.


\begin{figure*}
\begin{center}
\includegraphics[width=1.\textwidth]{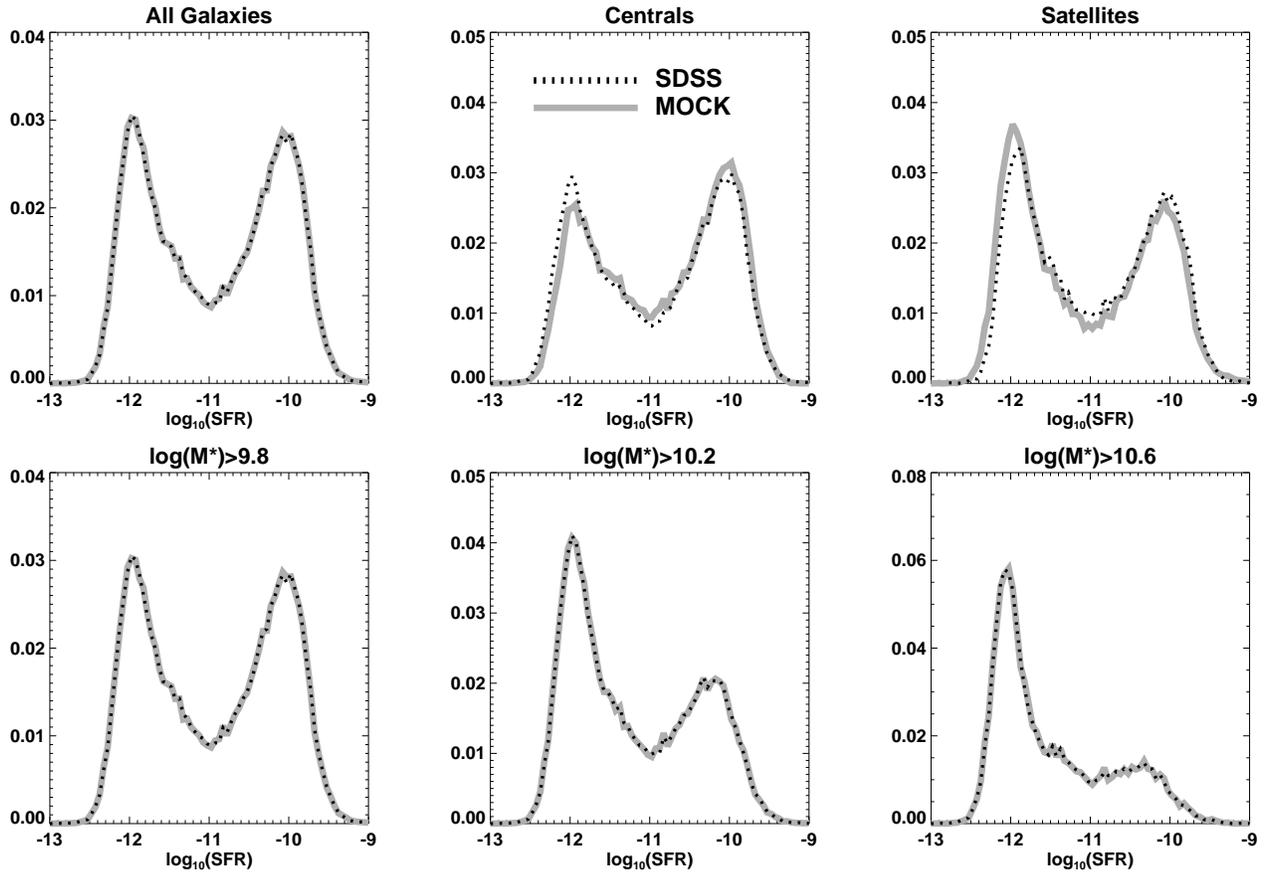}
\caption{ The probability distribution functions (PDFs) of the
  specific star formation rate (sSFR) of galaxies in our mock catalog
  (gray solid curves) as compared to those measured in the SDSS galaxy
  catalogs (dotted black curves).  By construction, the PDFs of sSFR
  of our mock galaxies are in exact agreement with the data for all
  the galaxies in our sample (top left panel) as well as three stellar
  mass threshold samples (bottom row):
  $\log_{10}(\Mstar)>[9.8,10.2,10.6]$.  Since our sSFR assignment to
  mock galaxies is blind to the distinction between central and
  satellite galaxies, the resultant PDFs  in the center and right
  panels of the top row are {\em predictions} of age matching, and
  demonstrate a non-trivial success of the technique.}
\label{fig:sm_SFR_PDFs}
\end{center}
\end{figure*}



\section{CATALOGS AND MEASUREMENTS}
\label{sec:data}

As our baseline galaxy sample and halo catalogs are identical to those
used in Papers I \& II, we only briefly sketch the essential  elements
of these catalogs here, and refer the reader to Papers I \& II for
further details. 

For our galaxy data, we use a volume-limited galaxy sample from DR7 of
the Sloan Digital Sky Survey \citep{york00a,DR7_09}, spanning the
redshift range $0.02<z<0.067,$ and complete to
$\log_{10}\Mstar/\Msun>9.8.$ We have identified galaxy groups in this
sample using the friends-of-friends algorithm presented in
\citet{berlind06}; for brevity, we refer to this sample as our
$\SMcat$ SDSS group catalog. 

We have cross-matched the $\SMcat$ catalog with the specific star
formation rate (sSFR) measurements taken from the MPA-JHU catalog,
publicly available at {\tt
  http://www.mpa-garching.mpg.de/SDSS/DR7}. The measurements are based
on the \citet{brinchmann_etal04} spectral reductions that
utilize the strength of $\mathrm{H}\alpha$ emission to estimate
present-day star formation activity, along with updated prescriptions
for fiber aperture corrections and active galactic nuclei (AGN) as
detailed in \citet{salim07}.  Specifically, sSFRs  are primarily
derived from emission lines (mostly $\mathrm{H}\alpha$), but in the
cases of strong AGN contamination or no measurable emission lines, the
sSFRs are inferred from Dn4000 in the galaxy spectrum \citep{kauffmann03a}.

Our mock catalog is constructed from halos and subhalos in the Bolshoi
$N-$body simulation \citep{bolshoi_11}, based on publicly available
ROCKSTAR merger trees and halo catalogs
\citep{rockstar_trees,rockstar}\footnote[1]{ROCKSTAR halo catalogs and
  merger trees are publicly available at {\tt
    http://hipacc.ucsc.edu/Bolshoi/MergerTrees.html}}. The simulation
has a volume of $250^{3}\,\hmpcvol$ with $2048^{3}$ dark matter
particles of mass $1.9\times10^{8}\hMsun,$ and a cold dark matter
($\Lambda$CDM) cosmological model with $\Omega_{\mathrm{m}}=0.27$,
$\Omega_{\Lambda}=0.73$, $\Omega_{\mathrm{b}}=0.042$, $h=0.7$,
$\sigma_{8}=0.82$.   For details on the Bolshoi database,
see \citet{riebe_etal13}.

From the $\SMcat$ galaxy sample, we present new measurements of the
two-point projected correlation function (2PCF), and the
galaxy--galaxy lensing signal ($\Delta\Sigma$), as a function of
stellar mass and SFR.  Specifically, we divide the galaxies into
`star-forming' and `quenched' populations by making a cut on the
measured value of sSFR at $10^{-11}\mathrm{yr}^{-1}.$  The clustering
and lensing measurements are performed both observationally and in the
simulation in the same manner as described in detail in Paper II.  We
make our mock galaxy catalog publicly available at {\tt
  http://logrus.uchicago.edu/$\sim$aphearin}.

In order to investigate satellite-specific properties of quenched and
star-forming galaxies addressed in \S~\ref{subsec:sat_results} we rely on our galaxy group finder to designate
central and satellite galaxies in both the mock and the SDSS data.
Specifically in each identified SDSS group, we label the galaxy with
the highest stellar mass as a central and all remaining galaxies in
the group as satellites. In our mock catalog, we follow the exact same
procedure. By reproducing the same
procedure in both SDSS data and mocks, we can compare the two without
concern for group finding errors in the central/satellite assignment.


\section{METHODOLOGY}
\label{sec:model}

Our main approach is to assign stellar masses and SFRs of galaxies to
(sub)halos within the Conditional Abundance Matching (CAM) formalism
(as fully detailed in \S~4.3 Paper II).  This formalism begins by
using the abundance matching technique \citep[e.g.,][]{kravtsov04a,vale_ostriker04,tasitsiomi_etal04,vale_ostriker06, trujillo_gomez11,rod_puebla12,watson_etal12b,hearin_etal12b,reddick12,kravtsov_smhm_14} to
assign stellar masses to halos and subhalos in Bolshoi to create a
volume-limited SDSS mock galaxy catalog. In particular, we abundance
match  the exact stellar mass function of our galaxy sample against
the (sub)halo property $\vpeak,$ (the highest circular velocity a halo
has had over its entire merger history)  using $\sim0.15$ dex of
scatter in $\Mstar$ at fixed $\vpeak,$ using the algorithm developed
in \citet{hearin_etal12b}.\footnote[2]{Although it has recently been
  shown that $\vpeak$ is typically set during a major merger and
  therefore unlikely  to be truly correlated with present day stellar
  mass in detail \citep{behroozi_etal13c}, the focus of the present
  paper  is on predicting present day SFR, and so we consider refining
  the traditional abundance matching algorithm  beyond our
  scope. However, we note that basing the stellar mass assignment on
  $\vacc$ (the maximum circular velocity of a halo when it accretes
  onto a larger halo, thus becoming a subhalo) rather than $\vpeak$
  has a negligible impact on the SFR predictions of the model.} Thus
as a result of this first phase of implementation of the CAM
technique,  our model naturally inherits the well known successes of
traditional abundance matching which has been shown to reproduce a variety
of observations including galaxy 2PCFs \citep{conroy06, reddick12},
close pair counts \citep{berrier06, berrier_cooke12}, $M_\ast-M_h$
relations \citep{conroy_wechsler09, wang_jing10, guo10,reddick12}, and
group multiplicity functions \citep{hearin_etal12b}.  We note that
\citet{kravtsov_smhm_14} recently demonstrated that improved
photometric techniques used to measure stellar mass
\citep{bernardi_etal13}  lead to quite a significant effect on the
stellar mass-to-halo mass relation predicted by abundance matching,
particularly for central galaxies of  halos at the high-mass end
($\Mhalo \gtrsim 10^{14} \hMsun$). We intend to explore the influence
of this systematic in future work, when we comprehensively explore the
age matching parameter space.

Once stellar masses have been assigned to our mock galaxies, we then
proceed to model galaxy SFRs using CAM, a general formalism to study
correlations at fixed mass between {\em any} galaxy property and {\em
  any} halo property.  The fundamental quantity in CAM is
$P(\Mstar,\xgal|\vmax,\xhalo),$ the probability that a galaxy of a
given stellar mass $\Mstar$ and galaxy property $\xgal$ resides in a
halo with circular velocity $\vmax$ and an additional halo property
$\xhalo$.  We choose the same specific implementation of CAM known as
{\em age matching}, introduced in Paper I and extended in Paper II.
In age matching, the quantity $\xgal$ is either $g-r$ color or sSFR,
and $\xhalo$ is the halo property $\zstarve$, which is characterized
by certain epochs in a halo's mass accretion history (MAH) presumed to
be linked to the starvation of the cold gas supply needed to continue
fueling star formation. These epochs include: the redshift a halo (1)
accretes onto a larger halo, (2) transitions from the fast- to
slow-accretion regimes (halo formation redshift $\zform$ for which we
use the concentration-based approximation introduced in
\citealt{wechsler02} and explain in detail in Appendix A of Paper I),
and (3) reaches a virial mass scale of $10^{12}\hMsun$.  Paper II
demonstrated that (1) and (2) are highly correlated, thus disregarding
halo accretion has proven to yield equally good model predictions.
Epoch (3) was introduced because the halo mass $10^{12}\hMsun$
demarcates a characteristic mass scale above which star formation is
believed to become rapidly inefficient due to AGN feedback
\citep{shankar_etal06,teyssier11,martizzi_etal12}, and to a lesser
extent due to  pressure-supported shocks that can heat infalling gas
to the virial temperature \citep{dekel_birnboim06}. A halo is assigned
a $\zstarve$ value based on whichever of these three epochs occurs
first in its MAH\footnote[3]{For details on calculating $\zstarve$
  from halo merger trees, see the appendix of Paper I.}, formally
written as $\zstarve\equiv
\mathrm{Max}\left\{\zacc,\zchar,\zform\right\}$. 

However, one may wonder whether correlating galaxy sSFR with
$\zstarve$ is necessary at all in the construction of a successful
model. This is due to the following chain of
logic: (1) more massive halos host galaxies of greater stellar mass,
(2) the quenched fraction increases with stellar mass, and therefore,
(3) quenched samples in a mock catalog constructed by drawing random
sSFRs without any $\zstarve$ ranking will preferentially weight higher
mass halos.  These higher mass halos have earlier assembly times (at
fixed stellar mass), and this so-called ``assembly bias'' has been
shown to affect the clustering of halos \citep[e.g.,][]{wechsler02}.
We have performed such a test (in Papers I \& II as well) and have found
that this effect alone is \emph{far} too weak to yield a working
model. This class of models that randomly draws sSFRs from the data
without any rank-ordering predicts minimal difference in the
clustering between the quenched and star-forming populations,
indicating that the effect of halo assembly bias alone is not a strong
enough mechanism to yield a significant enough split in the clustering
(see \citealt*{zentner_etal14} for a more comprehensive discussion).

In this paper, mock galaxies are assigned SFRs in the same manner as
they are assigned $g-r$ colors in Paper II: for a given fixed stellar
mass bin, sSFR values are drawn directly from the  probability
distribution function exhibited by our galaxy catalog, $\PsdssMstar.$
The collection of halos and subhalos in the corresponding stellar mass
bin are then rank-ordered by $\zstarve$, such that the most quenched
galaxy will be assigned to the halo with the largest $\zstarve$ value,
and so forth.  As seen in Fig.~5 of Paper II the halo formation epoch
$\zform$ (and hence, halo age) dominates the contribution to
$\zstarve$ over most of the stellar mass range probed by our
sample. We choose to not adopt a simpler model (i.e., $\zstarve = \zform$) for consistency in this trilogy of papers and we reserve any
model simplification/fine-tuning for future papers when we consider
additional statistics (e.g., galactic conformity) and push to other
stellar mass and redshift regimes.  In the end, age matching simply
posits that {\em quenched galaxies reside in old halos}, though the
more general CAM formalism allows for the exploration of any
galaxy-halo property correlation.

The above procedure results in a mock galaxy catalog whose sSFR
distribution is, by construction, in exact agreement with
$\PsdssMstar.$ This agreement is illustrated by the probability
distribution functions (PDFs) in the top left and bottom panels of
Fig.~\ref{fig:sm_SFR_PDFs}.  We emphasize that the purpose of the
rank-ordering is to introduce, at fixed stellar mass, a correlation
between galaxy SFR and $\zstarve$. However, as was the case for color
in Papers I $\&$ II, our technique only uses the property $\zstarve$
and $\PsdssMstar$ to assign SFRs to the mock galaxies, but does {\em
  not} distinguish between central and satellite galaxies in the SFR
assignment.  In fact, in age matching central and satellite galaxies
of the same stellar mass  have different SFR distributions strictly
due to differences in the assembly histories of host halos and
subhalos.  Therefore, there is no guarantee that our PDFs will be
correctly predicted for the sub-populations of centrals and
satellite. Nonetheless, as can be seen in the top middle and right
panels of Fig.~\ref{fig:sm_SFR_PDFs}, age matching does indeed
successfully predict central and satellite SFRs. We return to this
point in \S~\ref{sec:discussion} with the discussion of
Fig.~\ref{fig:age_cen_sat_quenching}.


\begin{figure*}
\begin{center}
\includegraphics[width=1.\textwidth]{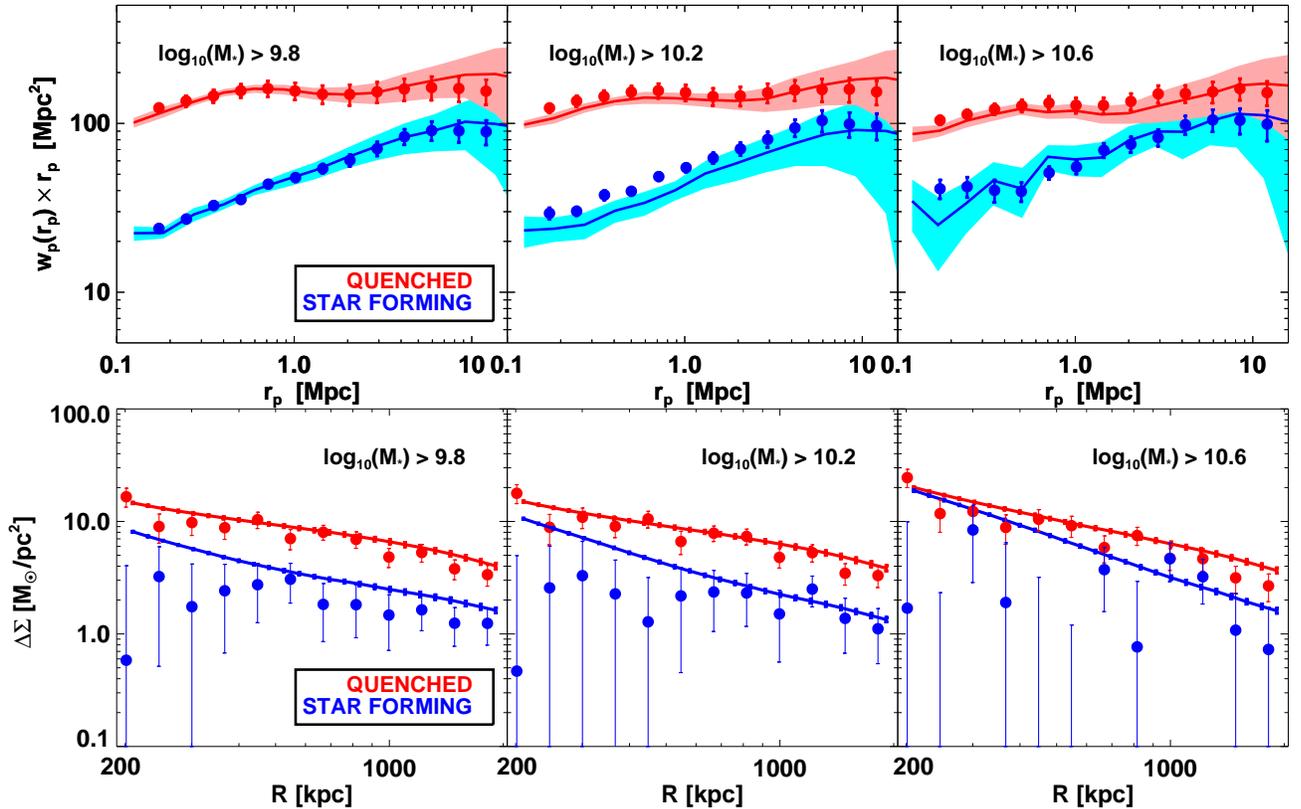}
\caption{SFR-dependent clustering and galaxy-galaxy lensing as a
  function of stellar mass as predicted by our age matching model
  versus new SDSS measurements. {\em {\bf Top Row:}} The projected
  correlation function (multiplied by $r_\mathrm{p}$) predicted by our
  model split into quenched and star-forming mock galaxy samples is
  shown with red and blue solid curves, respectively. Solid bands in
  each panel show the error in our model prediction estimated by
  jackknifing the octants of the simulation box.  Red (blue) points
  show our measurements of quenched (star-forming) SDSS galaxies
  (provided in Tables~\ref{tab:wp_thresholds_SF}
  $\&$~\ref{tab:wp_thresholds_QUENCHED}).  Errors on the measurements
  are computed from jackknife resampling of 50 equal-area regions on
  the sky.  {\em {\bf Bottom Row:}} Excess surface density
  $\Delta\Sigma$ as a function of stellar mass and SFR as predicted by
  our age matching model (red and blue solid curves solid curves) in
  comparison to new SDSS measurements. Our new SDSS $\Delta\Sigma$
  measurements are provided in Tables~3 $\&$ 4.  Errors on the SDSS
  lensing signal are derived by dividing the survey area into 200
  bootstrap subregions and generating 500 bootstrap-resampled
  datasets, while the age matching errors are computed via 27
  jackknife regions over the simulation volume.}
\label{fig:wp_gg}
\end{center}
\end{figure*}



\begin{figure*}
\begin{center}
\includegraphics[width=1.\textwidth]{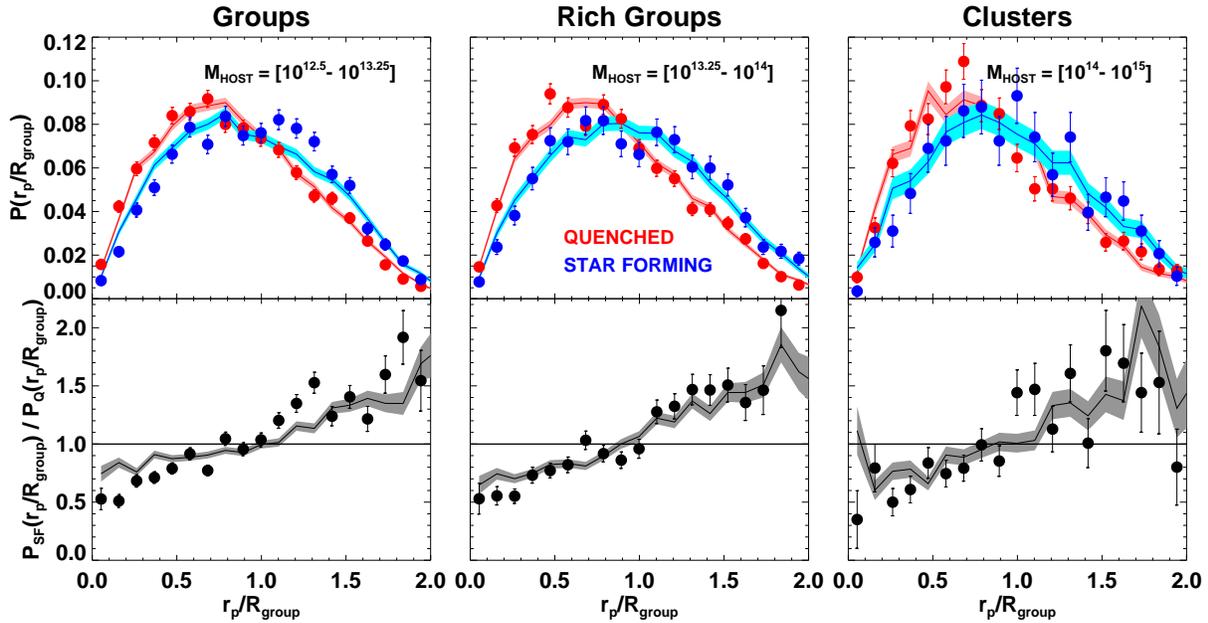}
\caption{PDFs of the radial distribution of quenched
  ($\mathrm{P_{\mathrm{Q}}(\mathrm{r_{p}/R_{group}})}$) and star
  forming ($\mathrm{P_{\mathrm{SF}}(\mathrm{r_{p}/R_{group}})}$)
  galaxies within and around groups, rich groups, and clusters as
  measured in our galaxy group catalog. {\em {\bf Top Row:}} Age
  matching predictions are shown as red and blue curves, respectively,
  versus the profiles measured in SDSS (red and blue filled circles).
  Three environmental regimes are considered corresponding to groups,
  rich groups, and clusters, which we define as having host halo
  masses of $10^{12.5-13.25}, 10^{13.25-14},$ and $10^{14-15} \hMsun$,
  respectively. The radial separation on the x-axis,
  $\mathrm{r_{p}/R_{group}}$, is defined as the projected separation,
  $\mathrm{r_{p}}$, divided by the rms group size,
  $\mathrm{R_{group}}$ ($\mathrm{R_{group}} \simeq 450 \hkpc$ for
  groups, $\simeq 650 \hkpc$ for rich groups, and $\simeq 1 \hmpc$ for
  clusters).  In each host halo mass regime, quenched galaxies are
  more centrally concentrated then their star-forming counter
  parts. {\em {\bf Bottom Row:}}  We divide the star-forming
  population PDF by that of the quenched population of the top panels
  to highlight differences between the quenched and star-forming
  radial profiles.  Results for the mock are shown as black solid
  lines with gray error bands, and filled black circles are for SDSS.
  Poisson errorbars are shown in all panels.}
\label{fig:rad_prof}
\end{center}
\end{figure*}



\begin{figure*}
\begin{center}
\includegraphics[width=1.\textwidth]{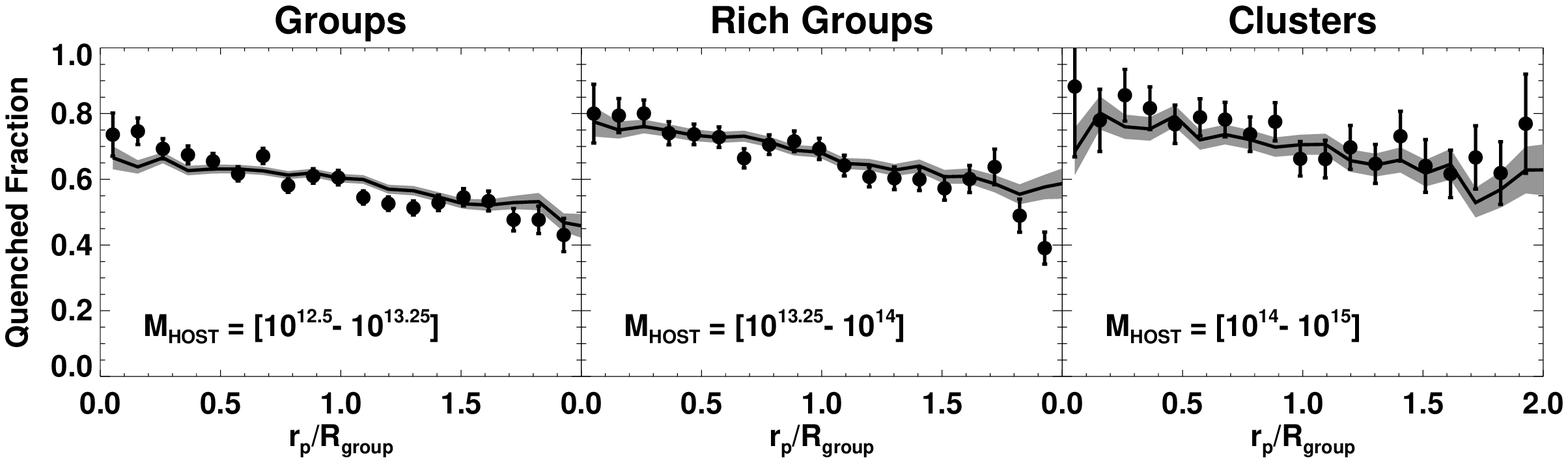}
\caption{The overall quenched fraction of satellite galaxies versus
  projected distance from the central galaxy within and around group-, rich group-, and
  cluster-mass halos.  The model prediction is striking: agreement
  with the SDSS data spans scales deep within host halos and extending out to
  radial separations well beyond the group radius for all three
  environments.  }
\label{fig:fquenched_rad_prof}
\end{center}
\end{figure*}



\section{Results}
\label{sec:results}

In this section, we present our main results.   We first demonstrate
in \S~\ref{subsec:2PCF_gg_results} that our age matching model
reproduces our new SDSS measurements of the projected galaxy 2PCF and
galaxy-galaxy lensing as a function of stellar mass, and divided into
quenched and star-forming populations.  We then focus squarely on
results pertinent to quenched and star-forming properties of satellite
galaxies as measured from the $\SMcat$ SDSS group catalog.
Specifically, in \S~\ref{subsec:sat_results} we compare our model
prediction for the radial distribution of quenched and star-forming
satellite galaxies within and around halos corresponding to group-,
rich group- and cluster-size halos. We then examine the radial
dependence of the quenched fraction of satellite galaxies in such
regimes.

\subsection{Galaxy Clustering and Galaxy-Galaxy Lensing}
\label{subsec:2PCF_gg_results}

\subsubsection{Clustering}
\label{subsubsec:wp}
We now investigate the success of age matching at predicting
SFR-dependent clustering.  Turning to the top row of
Fig.~\ref{fig:wp_gg}, red and blue solid curves are our model
predictions for the quenched and star-forming galaxy samples,
respectively.   Errors on the model $\wprp$ predictions are estimated
by jackknifing the octants of the simulation box. Red and blue filled
circles are new measurements from SDSS (see
Tables~\ref{tab:wp_thresholds_SF}
$\&$~\ref{tab:wp_thresholds_QUENCHED}).   Errors on the measurements
are computed from jackknife resampling of 50 equal-area regions on the
sky.  A detailed description for how we calculate the clustering in
the data and in the simulation can be found in \S~2.2 and \S~5.1 of
Paper II, respectively.  Our age matching predictions for the
SFR-dependent clustering are in excellent agreement with the data at
each stellar mass threshold sample and all projected separations.
However, as seen in the top, center panel of Fig.~2 of Paper II, there
is a slight under-prediction from abundance matching on small scales
($r_\mathrm{p} \lesssim 1 \hmpc$) for the $\mathrm{log}_{10}(\Mstar) >
10.2$ threshold sample.  This discrepancy naturally propagates through
to both the quenched and star-forming age matching predictions (top,
center panel), where the clustering amplitude on small scales is
suppressed with respect to the data. However, the {\em relative}
quenched and star-forming split of the model agrees well with that of
the data.

\subsubsection{Lensing}
\label{subsubsec:gg}
In addition to clustering, we also test our model against new SDSS
measurements of the SFR-dependent galaxy-galaxy lensing signal,
$\Delta\Sigma$, which are provided in Tables~3 $\&$ 4.  In Paper II we
describe how we calculate $\Delta\Sigma$ both in the data (\S~2.4) and
in the simulation (\S~5.2). As was shown in Fig.~2 of Paper II, we
accurately predict $\Delta\Sigma$ first at the abundance matching
level, though it should be noted that the amplitude of the abundance
matching prediction for each stellar mass threshold is slightly
boosted relative to the data. Thus we expect the age matching,
SFR-dependent $\Delta\Sigma$ amplitudes to be boosted for each sample
as well.  This is indeed the case as seen in the bottom row of
Fig.~\ref{fig:wp_gg}. 

SDSS data points for quenched and star-forming samples are represented
as red and blue filled circles, respectively, while red and blue solid
curves are the model predictions according to age matching.  Errors on
the $\Delta\Sigma$ model predictions are computed via 27 jackknife
regions over the Bolshoi simulation volume.  Errors on the SDSS
lensing signal are derived by dividing the survey area into 200
bootstrap subregions and generating 500 bootstrap-resampled datasets.
In light of the slight over-prediction of the model at the abundance
matching level, the {\em relative} separation in $\Delta\Sigma$
between quenched and star-forming samples is predicted reasonably well
for each stellar mass threshold, with the exception of the star
forming samples on the very smallest scales.

\subsection{Star-Forming and Quenched Satellite Galaxies within Galaxy Groups, Rich Groups, and Clusters}
\label{subsec:sat_results}

We now focus on results pertaining specifically to satellite galaxies.
By employing the same galaxy group finder to distinguish between
central and satellite galaxies in both our mock catalog and the SDSS
sample (see \S~\ref{sec:data} for details of the galaxy group finder
we employ), we investigate the radial distribution of quenched and
star-forming galaxies within group-, rich group- and cluster- size
halos and their surrounding larger scale environment.  We also study
the radial dependence of the satellite galaxy quenched fraction, and
test whether or not there is variation in the {\em slope} of the
profile for these three regimes.

\subsubsection{Radial Profiles of Star-Forming and Quenched Satellites}
\label{subsubsec:rad_profs}

In our study of the radial profiles of satellites, we consider three standard
regimes: groups, rich groups, and clusters, which we define as having
host halo masses of $10^{12.5-13.25}, 10^{13.25-14},$ and $10^{14-15}
\hMsun$, respectively.  Host halo masses are assigned to groups in the
traditional abundance matching manner, namely by matching the number
density of Bolshoi {\em host} halos rank-ordered by $\Mvir$ to the
number density of the groups rank-ordered by total stellar mass in the
group.  This procedure is done separately for the SDSS and mock
catalogs, for consistency.

For each satellite in both the age matching mock and SDSS data, we
measure $\mathrm{r_{p}}$,  the projected separation of the satellite
from the group's central galaxy. For each group,  we define
$\mathrm{R_{group}}$ to be the rms group size. An alternative choice
for group size  would be the virial radius presumed to be associated
with the group's halo, defined by  $\Mvir =
(4\pi/3)\Rvir^{3}\Delta_{\mathrm{vir}}\rho_{m},$ where $\rho_m$ is the
cosmic mean matter density, and  $\Delta_{\mathrm{vir}}\simeq360.$
However, $\Rvir$ and rms group size are in tight correspondence, with
a scatter of $\sim20\%$.  We find that $\mathrm{R_{group}} \simeq 450
\hkpc, 650 \hkpc,$ and $1 \hmpc$ for the group, rich group and cluster
regimes, respectively.

Using these measurements, in  the top row of Fig.~\ref{fig:rad_prof} we show the
PDFs of the radial distribution of quenched
($\mathrm{P_{\mathrm{Q}}(\mathrm{r_{p}/R_{group}})}$) and star-forming
galaxies ($\mathrm{P_{\mathrm{SF}}(\mathrm{r_{p}/R_{group}})}$). Model
predictions appear as red and blue curves, respectively,  SDSS
measurements appear as red and blue filled circles.  Poisson error
bars are shown in all panels. 

First note that in each panel it is clear that quenched galaxies are
more centrally concentrated than their star-forming counterparts.
There are in fact a larger {\em total} number of quenched than
star-forming satellites at all scales, but here the
$\mathrm{P_{\mathrm{SF}}(\mathrm{r_{p}/R_{group}})}$ and
$\mathrm{P_{\mathrm{Q}}(\mathrm{r_{p}/R_{group}})}$ are separately
unit-normalized in each host mass range,  thus highlighting the level
of accuracy of the age matching model with respect to the data. In the
bottom row of Fig.~\ref{fig:rad_prof} we divide
$\mathrm{P_{\mathrm{SF}}(\mathrm{r_{p}/R_{group}})}$ by
 $\mathrm{P_{\mathrm{Q}}(\mathrm{r_{p}/R_{group}})}$ to highlight
differences between the quenched and star-forming radial profiles.
Results for SDSS are shown as filled black circles and black
solid lines with gray error bands for the mock. Poisson errorbars are
shown in all panels.  Other than a slight discrepancy at very small
projected separations for the galaxy group halo mass scale, there is
excellent agreement between our model and these measurements for all
three host mass regimes, and at all projected separations.


\begin{figure}
  \includegraphics[width=.5\textwidth]{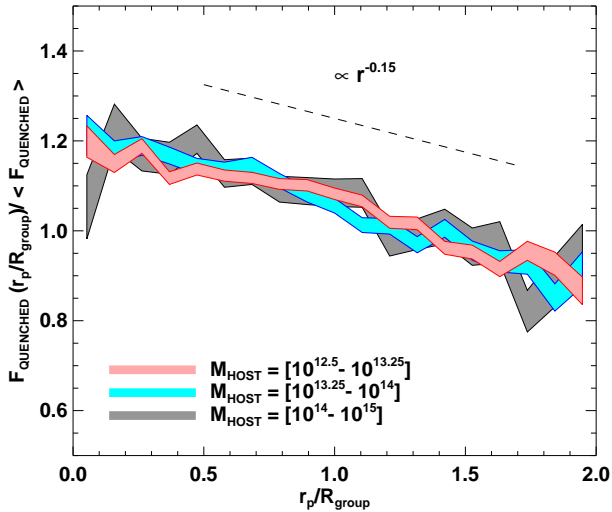}
\caption{The (lack of) environmental dependence of satellite quenching
  gradients. The model prediction in each panel of
  Fig.~\ref{fig:fquenched_rad_prof} is normalized by the {\em overall}
  mean satellite quenched fraction in each environment,  so that on
  the y-axis of this figure we plot
  $\mathrm{F_{QUENCHED}(\mathrm{r_{p}/R_{group}})/\langle F_{QUENCHED}
    \rangle}$.  What emerges is an $\sim \mathrm{r^{-0.15}}$ slope,
  {\em independent of environment}.} 
\label{fig:rad_prof_Mhost}
\end{figure}


\subsubsection{The Radial Dependence of the Satellite Galaxy Quenched Fraction}
\label{subsubsec:quenched_rad_profs}

Figure~\ref{fig:fquenched_rad_prof} shows the overall quenched
fraction as a function of $\mathrm{r_{p}/R_{group}}$.  In each bin of
$\mathrm{r_{p}/R_{group}},$  we compute the fraction of satellites
found in that bin that are quenched,
$\mathrm{F_{QUENCHED}}(\mathrm{r_{p}/R_{group}}).$  Again, the result
is compelling: the age matching prediction is in agreement with the
data on scales deep within the host halo of the group, and also extending out to scales well
beyond the group radii.  The overall amplitude of the radial quenched
fraction increases when going from the group to cluster regime, as has
been seen in other studies
\citep[e.g.,][]{wetzel_etal12b}. Additionally, one can notice by eye
that there is an apparent lack of $\Mhost-$dependence of the {\em
  slope} of the profiles.  To investigate this more closely,  we
separately normalize the radial quenched fractions for satellites in
each host halo mass range (i.e., each panel of
Fig.~\ref{fig:fquenched_rad_prof}) by the overall mean quenched
fraction of satellites in groups of that mass, given as
$\mathrm{F_{QUENCHED}(\mathrm{r_{p}/R_{group}})/\langle F_{QUENCHED}
  \rangle}$ on the y-axis of Fig.~\ref{fig:rad_prof_Mhost}.  The
result is an $\sim \mathrm{r^{-0.15}}$ slope for the radial quenched
fraction of satellite galaxies, {\em independent of environment}.

As  emphasized in Papers I $\&$ II, our age matching model has required
no parameter fitting to achieve the agreement between the predicted
and measured SFR-dependent galaxy statistics. There is good agreement
between our model and SDSS measurements for the predicted clustering,
lensing and satellite-specific spatial distributions within halos,
despite the fact that the SFR-halo assignment in age matching has {\em
  no explicit dependence on halo position or post-accretion orbital
  history}. Thus, in our mock the SFR-dependent spatial distributions
of both central and satellite galaxies simply emerges as a result of
the galaxy-halo co-evolution ansatz.


\section{Discussion $\&$ Interpretation}
\label{sec:discussion}

\subsection{Simplifying the Galaxy Evolution Picture with Age Matching}
\label{subsec:simplicity}

The primary result of this paper is that the simple age matching
model,  in which galaxies and halos co-evolve, such that quenched
galaxies reside in old halos, is able to predict a wide variety of
observed SDSS galaxy statistics for quenched and star-forming
galaxies.  In our model, there are  (1) no fine tuning or updates to
the age matching model that proved successful at reproducing
color-based SDSS measurements, (2) no distinction between central and
satellite galaxies when assigning SFRs, and (3) no explicit modeling
of post-accretion processes that are believed to stifle the star
formation in satellite galaxies (e.g., strangulation, ram pressure
stripping, etc.).

Let us consider these points in turn.  As discussed in
\S~\ref{sec:intro}, the color of a galaxy is known to be strongly
correlated with star formation activity.  For a variety of reasons,
though, the correspondence is not perfect.  For instance, active
galaxies can often be classified as red due to the presence of dust
\citep{maller_etal09,masters_etal10}. Color correlates with long term
mass accretion history in age matching because of the timescale
($\sim\Gyr$) to evolve from the blue to red sequence.  On the other
hand, the timescales relevant to, for example, H$\alpha$ indicators of
SFR are significantly shorter than timescales impacting color (e.g.,
$\sim10-100\Myr$ for the lifetime of O and B stars).  Therefore, it is
plausible that employing present day SFR in the age matching model may
not exhibit the same level of success as a model based on $g - r$
color.

We have shown that this is not the case: the SFR predictions of our
age matching implementation of CAM are equally successful as the
color-based predictions from Papers I $\&$ II. Certainly the
relatively tight scatter between $g-r$ and sSFR is partly responsible
for this dual success. In a follow-up paper to the present work
(Watson, Skibba $\&$ Hearin 2014, in prep), we will show that the
shortcomings of using broadband color as a proxy for present day star
formation activity have virtually no manifestation on the two-point
function. This surprising result, interesting in its own right,
provides further insight into the reason that our model is able
predict both $g-r$ color and SFR without modification.  The
explanation for this is simple: a star-forming galaxy appears red when
our line of sight to the galaxy lies in the plane of its disk; for a
{\em pair} of galaxies separated by $r\gtrsim100\kpc,$ the probability
of this occurrence is essentially independent due to the very weak
correlation between galaxy alignments \citep{zhang_etal13}.

\begin{figure*}
\begin{center}
  \includegraphics[width=1.\textwidth]{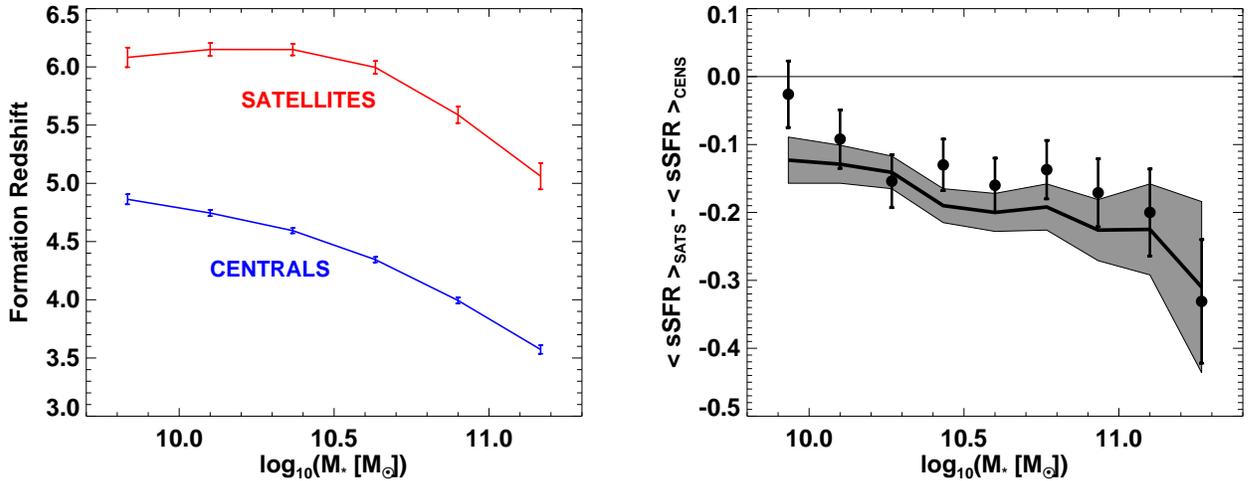}
\caption{{\bf Left Panel:} Formation epoch of central (blue curve) and
  satellite (red curve) galaxies as a function of stellar mass.
  Satellites in our model are more quenched than central galaxies of
  the same stellar mass simply because subhalos form earlier than host
  halos.  This fact about structure formation in CDM is what drives
  satellite quenching in the age matching model. {\bf Right Panel:}
  The difference between the average SFR of
  satellite and central galaxies as a function of stellar mass for
  SDSS (filled, black circles) and our age matching prediction (black
  solid line).  Poisson errors are shown for both data and the model.
  At fixed stellar mass satellites have lower SFRs than their central
  galaxy counterparts.}
\label{fig:age_cen_sat_quenching}
\end{center}
\end{figure*}

Point (2)  highlights the simplicity of age matching, as well as what
drives the satellite quenching predictions of the model.  Consider the
implications of the left panel of
Fig.~\ref{fig:age_cen_sat_quenching}.  We use our mock catalog to show
the average formation epoch of centrals (mock galaxies residing in
host halos) in comparison to satellites (subhalos).  As in Papers I
$\&$ II, we use the \citet{wechsler02} concentration-based definition
for the formation epoch of a halo. In age matching, despite there
being no distinction between central and satellite galaxies when
assigning a SFR, satellite galaxies are more quenched than their
central galaxy counterparts at fixed stellar mass simply because
subhalos form earlier than host halos. 

The empirical justification for this cornerstone of age matching is
illustrated in the right panel of
Fig.~\ref{fig:age_cen_sat_quenching}, where we show the difference in
the mean SFR of satellite and central galaxies in bins of fixed
stellar mass. For this figure, we now use the group-finder to identify
centrals and satellites, permitting a direct comparison to
observational data. SDSS measurements are shown as black, filled
circles and our age matching model prediction is shown as the solid
black line with a gray error band (all errors are Poisson).  At fixed
stellar mass, satellites are more quenched than centrals in both the
data and the model, and the observed quenching difference is
quantitatively consistent with the difference implied by the relative
formation times of host halos and subhalos.

This observation is closely connected to point (3).  Age matching does
not require any explicit modeling of post-infall effects on satellite
galaxy quenching.  The virial radius $\Rvir$ of host halos only enters
into our model through the definition of $\zacc,$ the epoch when a
halo accretes onto a larger halo, thus becoming a subhalo.  However,
recall that in age matching, SFR is determined by $\zstarve$, the
redshift in a (sub)halo's MAH when it is deemed to be starved of the
cold gas supply needed to continue fueling star formation.  Formally,
$\zstarve\equiv \mathrm{Max}\left\{\zacc,\zchar,\zform\right\}$, and
as we showed in Paper II, the epoch $\zacc$ has an essentially
negligible impact on $\zstarve$ at all stellar masses, a result which
also holds true in the present work. {\em Thus in our model, $\Rvir$
  does not mark a special transition region in the evolution of a
  satellite,} and yet we accurately predict the radial profiles of
quenched and star-forming galaxies both inside and well beyond the
group radius, as well as the so-called  ``excess quenched fraction''
of satellites (right panel of Fig.~\ref{fig:age_cen_sat_quenching}).


This qualitatively distinguishes age matching from conventional
semi-analytic and empirical models of satellite quenching. We note,
however, that the lack of explicit appearance of $\Rvir$ in our model
does {\em not} imply that post-accretion processes are necessarily
irrelevant to satellite quenching, since there is a significant
correlation between the time a subhalo accretes onto a larger halo and
the time the subhalo formed (see Fig.~6 of Paper II), rather that the
overallin fluence of post-accretion physics has been overstated in the
literature. Nonetheless, we will show in a pair of companion papers to
this one that recent observations of SFR trends in the low-redshift
universe {\em do} favor a scenario in which quenching is impacted by
physical processes that operate on scales far larger than $\Rvir,$
{\em even for central galaxies}, as discussed in the following
section. 

\subsection{Discriminating between Competing Quenching Models}
\label{subsec:conformity}

As discussed in detail in \citet*{zentner_etal14}, the success of age
matching has exposed fundamental degeneracies in traditional
approaches to galaxy-halo modeling such as the Halo Occupation
Distribution \citep[HOD,
  e.g.,][]{seljak00,cooray02,berlind02,berlind03,zheng05,skibba_sheth09,watson_etal10,watson_etal12a}
and Conditional Luminosity Function \citep[CLF,
  e.g.,][]{yang03,vdBosch13}.  While it is true that age matching is
based on $\vmax$ to set the stellar mass or luminosity content of
halos, the assignment of the additional galaxy properties of color or
SFR is based on halo assembly history.  Conversely, HOD modeling
beyond just stellar mass- or luminosity - dependent clustering,
i.e. the color or SFR dependence
\citep[e.g.,][]{zehavi05a,zehavi11,skibba09a,tinker_etal13,guo_SDSS14},
is still exclusively governed by $\Mvir$ and no other halo
property. And yet, both classes of models give very good descriptions
of a wide variety of measurements of the galaxy distribution.

These considerations apply equally well to degeneracies with
other common models of galaxy evolution. Indeed, both HODs and CLFs
enjoy comparable levels of qualitative successes in reproducing
observed statistics such as those presented in this work and the
previous age matching papers. There is thus some legitimate cause for
concern that conventional statistics describing the galaxy
distribution are inadequate to conclusively discriminate between
competing models.  One particularly interesting
measurement is that of {\em galactic conformity}, a feature in the
galaxy distribution first discovered by \citet{weinmann06b}. In their
analysis of an SDSS galaxy group catalog, \citet{weinmann06b} showed
that in group systems of the same halo mass, satellites in groups with
a red central tend to be redder than satellites in groups with a blue
central. In another recently reported detection of conformity,
\citet{kauffmann_etal13} showed that in an SDSS sample of central
galaxies of the same stellar mass, the environment surrounding
quenched central galaxies exhibits, on average, an attenuated SFR
relative to the environment around star-forming centrals, a
correlation that persists out to scales of $R\sim5\Mpc,$ far outside
the virial radius of the host halo of the centrals.

As we show in a recent paper \citep*{APH_DFW_vdB}, these closely
related signals are formally distinct in the following sense: the
\citet{weinmann06b} notion of conformity pertains to SFR correlations
between central and satellite galaxies in the same dark matter halo,
while the larger scale \citet{kauffmann_etal13} signal pertains to SFR
correlations in distinct halos. We contend that no galaxy evolution
model in which central galaxy SFR is exclusively determined by halo
mass $\Mvir$ (and subsequently virial radius $\Rvir$) can account for either signal, as there would be no
mechanism by which such correlations could arise. However, in age
matching, galaxies in the same  environment evolve from collapsed
peaks of the same region of the initial cosmic density field. Thus the
known correlation between the formation times of nearby halos
\citep[e.g.,][]{sheth_tormen04,wechsler06} naturally gives rise to
correlated stellar mass assembly histories of nearby galaxies.  In
\citet*{APH_DFW_vdB} we demonstrate that age matching predicts
galactic conformity of both the \citet{weinmann06b} and
\citet{kauffmann_etal13} varieties, with no modifications to the model
presented in this work. Since it was shown in \citet{kauffmann_etal13}
that the \citet{guo_etal11} semi-analytic model (SAM) does not predict
conformity, this signal appears to be a promising testbed for the
further development of galaxy evolution models.\footnote[5]{For
  example,  the ``pre-heating'' of gas in the inter-galactic medium
  implemented in the SAM recently introduced  in \citet{lu_etal14} is
  a promising mechanism by which conformity may arise, as discussed in
  \citet{kauffmann_etal13}.}

\subsection{Future Directions}
\label{subsec:future}

This trilogy of age matching papers has revealed that there is a
surprisingly simple relationship  between the star formation activity
of a galaxy and the assembly history of its dark matter halo.
However, there are two clear paths to challenging the age matching
picture of galaxy and halo co-evolution.  First, our model has only
been tested for central and satellite galaxies of
$\log_{10}(\Mstar)>9.8$.  However, using observations of classical
dwarf galaxies in SDSS, \citet{geha_etal12} discovered that there is
an apparent stellar mass threshold of $\log_{10}(\Mstar)=9.0,$ below
which quenched galaxies do not exist in the field.  In a recent study
of this SDSS dwarf sample, \citet{wheeler_etal14}
demonstrated that the so-called ``quenching timescale'' after a
satellite first crosses the virial radius $\Rvir$ of its host halo
must be implausibly long ($\gtrsim 9 \mathrm{Gyr}$) to produce the
trends reported in \citet{geha_etal12}.  These results are
intriguingly in keeping with the notion supported by age matching that
the role of $\Rvir$ has been over-estimated in the literature.  In
future work, we aim to apply the CAM modeling technique to dwarf
galaxy samples to investigate whether the SFR  trends exhibited  by
galaxies in this mass range are also reflected in a simple way by the
evolutionary history of dark matter halos.

The second consideration will be confronting age matching with
observations at higher redshift. We will soon take this crucial next
step thanks to high-completeness data sets such as PRIMUS
\citep{Primus_Coil}, GAMA \citep{GAMA_Driver} and VIPERS
\citep{VIPERS_Guzzo}. 

Ultimately, the power of this class of semi-empirical models is their
ability to be used as ``training sets'' to help inform more
complicated models of galaxy formation that include prescriptions for
physical processes \emph{a priori}.  Specifically, SAMs and
hydrodynamic simulations can draw insight from age matching in order
to improve their input physical recipes.

\section{Summary}
\label{sec:summary}

In this paper we have explored the hypothesis that the star formation
rates (SFRs) of galaxies can be determined via the ansatz that
galaxies co-evolve with their dark matter halos. Specifically, we have
studied the age matching formalism introduced in \citet{HW13a}, whose
central tenet is that red/quenched galaxies reside in old halos, and
conversely for blue/star-forming galaxies.  This simple formalism has
been proven to be remarkably powerful, yielding accurate predictions
of SDSS color-dependent clustering and galaxy-galaxy lensing, as well
as a variety of galaxy group-based statistics.  In this work have
confronted our age matching formalism with SFR-dependent, low-redshift
galaxy statistics.  Specifically, we have found the following
principal results.

\begin{itemize}
\item{We present new measurements of SDSS clustering and galaxy-galaxy
  lensing as a function of stellar mass, split into distinct quenched
  and star-forming populations.  The same age matching prescription
  introduced in Paper I and extended in Paper II adapts seamlessly to
  accurately predict these SFR-dependent observations, {\em without
    necessitating updates to the model or fine-tuning/fitting of
    parameters}.}
\item{Age matching predictions are in excellent
  agreement with the observed radial distribution of star-forming and quenched
  satellite galaxies within and around galaxy group, rich group, and cluster
  environments, a success that extends significantly beyond the group radius.}
\item{We demonstrate the lack of halo mass-dependence in the slope
  of the radial quenched fraction of satellites, finding an $\sim
  \mathrm{r}^{-.15}$ gradient {\em independent of environment}.}
\item{We make our mock galaxy catalog publicly available at {\tt http://logrus.uchicago.edu/$\sim$aphearin}.}  
\end{itemize}

These findings provide compelling evidence for the co-evolution of
halos and galaxies, and are highly suggestive of the conclusion that
the existing literature has over-estimated the role of post-accretion
processes on attenuating star formation in satellite galaxies. We
consider the myriad successes of our model to indicate that there does
indeed exist a simple relation between  cosmic star formation history
of galaxies and the dark side of the universe.


\section*{acknowledgments}

We would like to thank Andrey Kravtsov for productive discussions and
the anonymous referee for many insightful recommendations to improve
the manuscript. We would also like to thank John Fahey for {\em The
  Great Santa Barbara Oil Slick.}   DFW is supported by the National
Science Foundation under Award No. AST-1202698.  APH supported by the
U.S. Department of Energy under contract No. DE-AC02-07CH11359.  RAS
is supported by the NSF grant AST-1055081.  ARZ is supported by the
U. S. National Science Foundation through grant AST 1108802 and by the
University of Pittsburgh.  MRB was supported in part by the Kavli
Institute for Cosmological Physics at the University of Chicago
through grant NSF PHY-1125897 and an endowment from the Kavli
Foundation and its founder Fred Kavli.  AAB is supported by NSF grant
AST-1109789.  A portion of this work was also supported by the
National Science Foundation under grant PHYS-1066293 and the
hospitality of the Aspen Center for Physics.  PSB was supported by a
Giacconi Fellowship through the Space Telescope Science Institute,
which is operated by AURA for NASA under contract NAS5-26555.  This
work made extensive use of the NASA Astrophysics Data System and the
\verb+arxiv.org+ preprint server.  


\bibliography{./citations.bib}

\begin{thebibliography}{}

\bibitem[\protect\citeauthoryear{{Abazajian}, {Adelman-McCarthy},
  {Ag{\"u}eros}, {Allam}, {Allende Prieto}, {An}, {Anderson}, {Anderson},
  {Annis}, {Bahcall} \& et al.}{{Abazajian} et~al.}{2009}]{DR7_09}
{Abazajian} K.~N.,  {Adelman-McCarthy} J.~K.,  {Ag{\"u}eros} M.~A.,  {Allam}
  S.~S.,  {Allende Prieto} C.,  {An} D.,  {Anderson} K.~S.~J.,  {Anderson}
  S.~F.,  {Annis} J.,  {Bahcall} N.~A.,    et al. 2009, \apjs, 182, 543

\bibitem[\protect\citeauthoryear{{Baldry}, {Glazebrook}, {Brinkmann},
  {Ivezi{\'c}}, {Lupton}, {Nichol} \& {Szalay}}{{Baldry}
  et~al.}{2004}]{baldry04}
{Baldry} I.~K.,  {Glazebrook} K.,  {Brinkmann} J.,  {Ivezi{\'c}} {\v Z}.,
  {Lupton} R.~H.,  {Nichol} R.~C.,    {Szalay} A.~S.,  2004, \apj, 600, 681

\bibitem[\protect\citeauthoryear{{Balogh}, {Morris}, {Yee}, {Carlberg} \&
  {Ellingson}}{{Balogh} et~al.}{1999}]{balogh99}
{Balogh} M.~L.,  {Morris} S.~L.,  {Yee} H.~K.~C.,  {Carlberg} R.~G.,
  {Ellingson} E.,  1999, \apj, 527, 54

\bibitem[\protect\citeauthoryear{{Behroozi} et~al.,}{{Behroozi}
  et~al.}{2013a}]{rockstar_trees}
{Behroozi} P.~S.,  et~al., 2013a, \apj, 763, 18

\bibitem[\protect\citeauthoryear{{Behroozi} et~al.,}{{Behroozi}
  et~al.}{2013b}]{rockstar}
{Behroozi} P.~S.,  et~al., 2013b, \apj, 762, 109

\bibitem[\protect\citeauthoryear{{Behroozi}, {Wechsler} \& {Conroy}}{{Behroozi}
  et~al.}{2013a}]{behroozi13b}
{Behroozi} P.~S.,  {Wechsler} R.~H.,    {Conroy} C.,  2013a, \apjl, 762, L31

\bibitem[\protect\citeauthoryear{{Behroozi}, {Wechsler} \& {Conroy}}{{Behroozi}
  et~al.}{2013b}]{behroozi13}
{Behroozi} P.~S.,  {Wechsler} R.~H.,    {Conroy} C.,  2013b, \apj, 770, 57

\bibitem[\protect\citeauthoryear{{Behroozi}, {Wechsler}, {Lu}, {Hahn}, {Busha},
  {Klypin} \& {Primack}}{{Behroozi} et~al.}{2013}]{behroozi_etal13c}
{Behroozi} P.~S.,  {Wechsler} R.~H.,  {Lu} Y.,  {Hahn} O.,  {Busha} M.~T.,
  {Klypin} A.,    {Primack} J.~R.,  2013, ArXiv:1310.2239

\bibitem[\protect\citeauthoryear{{Bell} et~al.,}{{Bell}  et~al.}{2004}]{bell04}
{Bell} E.~F.,  et~al., 2004, \apj, 608, 752

\bibitem[\protect\citeauthoryear{{Berlind} et~al.,}{{Berlind}
  et~al.}{2003}]{berlind03}
{Berlind} A.~A.,  et~al., 2003, \apj, 593, 1

\bibitem[\protect\citeauthoryear{{Berlind} et~al.,}{{Berlind}
  et~al.}{2006}]{berlind06}
{Berlind} A.~A.,  et~al., 2006, \apjs, 167, 1

\bibitem[\protect\citeauthoryear{{Berlind} \& {Weinberg}}{{Berlind} \&
  {Weinberg}}{2002}]{berlind02}
{Berlind} A.~A.,  {Weinberg} D.~H.,  2002, \apj, 575, 587

\bibitem[\protect\citeauthoryear{{Bernardi}, {Meert}, {Sheth}, {Vikram},
  {Huertas-Company}, {Mei} \& {Shankar}}{{Bernardi}
  et~al.}{2013}]{bernardi_etal13}
{Bernardi} M.,  {Meert} A.,  {Sheth} R.~K.,  {Vikram} V.,  {Huertas-Company}
  M.,  {Mei} S.,    {Shankar} F.,  2013, \mnras, 436, 697

\bibitem[\protect\citeauthoryear{{Berrier}, {Bullock}, {Barton}, {Guenther},
  {Zentner} \& {Wechsler}}{{Berrier} et~al.}{2006}]{berrier06}
{Berrier} J.~C.,  {Bullock} J.~S.,  {Barton} E.~J.,  {Guenther} H.~D.,
  {Zentner} A.~R.,    {Wechsler} R.~H.,  2006, \apj, 652, 56

\bibitem[\protect\citeauthoryear{{Berrier} \& {Cooke}}{{Berrier} \&
  {Cooke}}{2012}]{berrier_cooke12}
{Berrier} J.~C.,  {Cooke} J.,  2012, \mnras, 426, 1647

\bibitem[\protect\citeauthoryear{{Blanton}, {Eisenstein}, {Hogg}, {Schlegel} \&
  {Brinkmann}}{{Blanton} et~al.}{2005}]{blanton05}
{Blanton} M.~R.,  {Eisenstein} D.,  {Hogg} D.~W.,  {Schlegel} D.~J.,
  {Brinkmann} J.,  2005, \apj, 629, 143

\bibitem[\protect\citeauthoryear{{Blanton} et~al.,}{{Blanton}
  et~al.}{2003}]{blanton03}
{Blanton} M.~R.,  et~al., 2003, \apj, 594, 186

\bibitem[\protect\citeauthoryear{{Brinchmann}, {Charlot}, {White}, {Tremonti},
  {Kauffmann}, {Heckman} \& {Brinkmann}}{{Brinchmann}
  et~al.}{2004}]{brinchmann_etal04}
{Brinchmann} J.,  {Charlot} S.,  {White} S.~D.~M.,  {Tremonti} C.,  {Kauffmann}
  G.,  {Heckman} T.,    {Brinkmann} J.,  2004, \mnras, 351, 1151

\bibitem[\protect\citeauthoryear{{Carollo} et~al.,}{{Carollo}
  et~al.}{2013}]{carollo_etal12}
{Carollo} C.~M.,  et~al., 2013, \apj, 776, 71

\bibitem[\protect\citeauthoryear{{Coil} et~al.,}{{Coil}
  et~al.}{2011}]{Primus_Coil}
{Coil} A.~L.,  et~al., 2011, \apj, 741, 8

\bibitem[\protect\citeauthoryear{{Conroy} \& {Wechsler}}{{Conroy} \&
  {Wechsler}}{2009}]{conroy_wechsler09}
{Conroy} C.,  {Wechsler} R.~H.,  2009, \apj, 696, 620

\bibitem[\protect\citeauthoryear{{Conroy}, {Wechsler} \& {Kravtsov}}{{Conroy}
  et~al.}{2006}]{conroy06}
{Conroy} C.,  {Wechsler} R.~H.,    {Kravtsov} A.~V.,  2006, \apj, 647, 201

\bibitem[\protect\citeauthoryear{{Cooper} et~al.,}{{Cooper}
  et~al.}{2006}]{cooper06}
{Cooper} M.~C.,  et~al., 2006, \mnras, 370, 198

\bibitem[\protect\citeauthoryear{{Cooper} et~al.,}{{Cooper}
  et~al.}{2012}]{cooper12}
{Cooper} M.~C.,  et~al., 2012, \mnras, 419, 3018

\bibitem[\protect\citeauthoryear{{Cooray} \& {Sheth}}{{Cooray} \&
  {Sheth}}{2002}]{cooray02}
{Cooray} A.,  {Sheth} R.,  2002, \physrep, 372, 1

\bibitem[\protect\citeauthoryear{{Dekel} \& {Birnboim}}{{Dekel} \&
  {Birnboim}}{2006}]{dekel_birnboim06}
{Dekel} A.,  {Birnboim} Y.,  2006, \mnras, 368, 2

\bibitem[\protect\citeauthoryear{{Driver} et~al.,}{{Driver}
  et~al.}{2011}]{GAMA_Driver}
{Driver} S.~P.,  et~al., 2011, \mnras, 413, 971

\bibitem[\protect\citeauthoryear{{Geha}, {Blanton}, {Yan} \& {Tinker}}{{Geha}
  et~al.}{2012}]{geha_etal12}
{Geha} M.,  {Blanton} M.~R.,  {Yan} R.,    {Tinker} J.~L.,  2012, \apj, 757, 85

\bibitem[\protect\citeauthoryear{{Gunn} \& {Gott} III}{{Gunn} \&
  {Gott}}{1972}]{gunn_gott72}
{Gunn} J.~E.,  {Gott} III J.~R.,  1972, \apj, 176, 1

\bibitem[\protect\citeauthoryear{{Guo} et~al.,}{{Guo}
  et~al.}{2014a}]{guo_etal14}
{Guo} H.,  et~al., 2014a, ArXiv:1401.3009

\bibitem[\protect\citeauthoryear{{Guo} et~al.,}{{Guo}
  et~al.}{2014b}]{guo_SDSS14}
{Guo} H.,  et~al., 2014b, ArXiv: 1401.3009

\bibitem[\protect\citeauthoryear{{Guo}, {Cole}, {Eke} \& {Frenk}}{{Guo}
  et~al.}{2011}]{guo_etal11}
{Guo} Q.,  {Cole} S.,  {Eke} V.,    {Frenk} C.,  2011, \mnras, 417, 370

\bibitem[\protect\citeauthoryear{{Guo}, {White}, {Li} \&
  {Boylan-Kolchin}}{{Guo} et~al.}{2010}]{guo10}
{Guo} Q.,  {White} S.,  {Li} C.,    {Boylan-Kolchin} M.,  2010, \mnras, 404,
  1111

\bibitem[\protect\citeauthoryear{{Guzzo} et~al.,}{{Guzzo}
  et~al.}{2013}]{VIPERS_Guzzo}
{Guzzo} L.,  et~al., 2013, ArXiv:1303.2623

\bibitem[\protect\citeauthoryear{{Hearin} et~al.,}{{Hearin}
  et~al.}{2014}]{HW13b}
{Hearin} A.~P.,  et~al., 2014, \mnras, 444, 729

\bibitem[\protect\citeauthoryear{{Hearin} \& {Watson}}{{Hearin} \&
  {Watson}}{2013}]{HW13a}
{Hearin} A.~P.,  {Watson} D.~F.,  2013, \mnras, 435, 1313

\bibitem[\protect\citeauthoryear{{Hearin}, {Watson} \& {van den
  Bosch}}{{Hearin} et~al.}{2014}]{APH_DFW_vdB}
{Hearin} A.~P.,  {Watson} D.~F.,    {van den Bosch} F.~C.,  2014,
  ArXiv:1404.6524

\bibitem[\protect\citeauthoryear{{Hearin}, {Zentner}, {Berlind} \&
  {Newman}}{{Hearin} et~al.}{2013}]{hearin_etal12b}
{Hearin} A.~P.,  {Zentner} A.~R.,  {Berlind} A.~A.,    {Newman} J.~A.,  2013,
  \mnras, 433, 659

\bibitem[\protect\citeauthoryear{{Kauffmann} et~al.,}{{Kauffmann}
  et~al.}{2003}]{kauffmann03a}
{Kauffmann} G.,  et~al., 2003, \mnras, 341, 33

\bibitem[\protect\citeauthoryear{{Kauffmann}, {Li}, {Zhang} \&
  {Weinmann}}{{Kauffmann} et~al.}{2013}]{kauffmann_etal13}
{Kauffmann} G.,  {Li} C.,  {Zhang} W.,    {Weinmann} S.,  2013, \mnras, 430,
  1447

\bibitem[\protect\citeauthoryear{{Klypin}, {Trujillo-Gomez} \&
  {Primack}}{{Klypin} et~al.}{2011}]{bolshoi_11}
{Klypin} A.~A.,  {Trujillo-Gomez} S.,    {Primack} J.,  2011, \apj, 740, 102

\bibitem[\protect\citeauthoryear{{Kravtsov}, {Vikhlinin} \&
  {Meshscheryakov}}{{Kravtsov} et~al.}{2014}]{kravtsov_smhm_14}
{Kravtsov} A.,  {Vikhlinin} A.,    {Meshscheryakov} A.,  2014, ArXiv: 1401.7329

\bibitem[\protect\citeauthoryear{{Kravtsov}}{{Kravtsov}}{2013}]{kravtsov_size_%
Rvir13}
{Kravtsov} A.~V.,  2013, \apjl, 764, L31

\bibitem[\protect\citeauthoryear{{Kravtsov}, {Berlind}, {Wechsler}, {Klypin},
  {Gottl{\"o}ber}, {Allgood} \& {Primack}}{{Kravtsov}
  et~al.}{2004}]{kravtsov04a}
{Kravtsov} A.~V.,  {Berlind} A.~A.,  {Wechsler} R.~H.,  {Klypin} A.~A.,
  {Gottl{\"o}ber} S.,  {Allgood} B.,    {Primack} J.~R.,  2004, \apj, 609, 35

\bibitem[\protect\citeauthoryear{{Larson}, {Tinsley} \& {Caldwell}}{{Larson}
  et~al.}{1980}]{larson80}
{Larson} R.~B.,  {Tinsley} B.~M.,    {Caldwell} C.~N.,  1980, \apj, 237, 692

\bibitem[\protect\citeauthoryear{{Leitner}}{{Leitner}}{2012}]{leitner12}
{Leitner} S.~N.,  2012, \apj, 745, 149

\bibitem[\protect\citeauthoryear{{Li}, {Kauffmann}, {Jing}, {White},
  {B{\"o}rner} \& {Cheng}}{{Li} et~al.}{2006}]{li06}
{Li} C.,  {Kauffmann} G.,  {Jing} Y.~P.,  {White} S.~D.~M.,  {B{\"o}rner} G.,
   {Cheng} F.~Z.,  2006, \mnras, 368, 21

\bibitem[\protect\citeauthoryear{{Lu}, {Mo}, {Lu}, {Katz}, {Weinberg}, {van den
  Bosch} \& {Yang}}{{Lu} et~al.}{2014}]{lu_etal14}
{Lu} Z.,  {Mo} H.~J.,  {Lu} Y.,  {Katz} N.,  {Weinberg} M.~D.,  {van den Bosch}
  F.~C.,    {Yang} X.,  2014, \mnras

\bibitem[\protect\citeauthoryear{{Maller}, {Berlind}, {Blanton} \&
  {Hogg}}{{Maller} et~al.}{2009}]{maller_etal09}
{Maller} A.~H.,  {Berlind} A.~A.,  {Blanton} M.~R.,    {Hogg} D.~W.,  2009,
  \apj, 691, 394

\bibitem[\protect\citeauthoryear{{Martizzi}, {Teyssier} \& {Moore}}{{Martizzi}
  et~al.}{2012}]{martizzi_etal12}
{Martizzi} D.,  {Teyssier} R.,    {Moore} B.,  2012, \mnras, 420, 2859

\bibitem[\protect\citeauthoryear{{Masters} et~al.,}{{Masters}
  et~al.}{2010}]{masters_etal10}
{Masters} K.~L.,  et~al., 2010, \mnras, 404, 792

\bibitem[\protect\citeauthoryear{{Moore}, {Lake} \& {Katz}}{{Moore}
  et~al.}{1998}]{moore_etal98}
{Moore} B.,  {Lake} G.,    {Katz} N.,  1998, \apj, 495, 139

\bibitem[\protect\citeauthoryear{{Mostek}, {Coil}, {Cooper}, {Davis}, {Newman}
  \& {Weiner}}{{Mostek} et~al.}{2013}]{mostek12}
{Mostek} N.,  {Coil} A.~L.,  {Cooper} M.,  {Davis} M.,  {Newman} J.~A.,
  {Weiner} B.~J.,  2013, \apj, 767, 89

\bibitem[\protect\citeauthoryear{{Moster}, {Naab} \& {White}}{{Moster}
  et~al.}{2013}]{moster13}
{Moster} B.~P.,  {Naab} T.,    {White} S.~D.~M.,  2013, \mnras, 428, 3121

\bibitem[\protect\citeauthoryear{{Norberg} et~al.,}{{Norberg}
  et~al.}{2002}]{norberg02}
{Norberg} P.,  et~al., 2002, \mnras, 332, 827

\bibitem[\protect\citeauthoryear{{Peng} et~al.,}{{Peng}
  et~al.}{2010}]{peng_etal10}
{Peng} Y.-j.,  et~al., 2010, \apj, 721, 193

\bibitem[\protect\citeauthoryear{{Peng}, {Lilly}, {Renzini} \&
  {Carollo}}{{Peng} et~al.}{2012}]{peng_etal12}
{Peng} Y.-j.,  {Lilly} S.~J.,  {Renzini} A.,    {Carollo} M.,  2012, \apj, 757,
  4

\bibitem[\protect\citeauthoryear{{Purcell}, {Bullock} \& {Zentner}}{{Purcell}
  et~al.}{2007}]{purcell_etal07}
{Purcell} C.~W.,  {Bullock} J.~S.,    {Zentner} A.~R.,  2007, \apj, 666, 20

\bibitem[\protect\citeauthoryear{{Reddick}, {Wechsler}, {Tinker} \&
  {Behroozi}}{{Reddick} et~al.}{2013}]{reddick12}
{Reddick} R.~M.,  {Wechsler} R.~H.,  {Tinker} J.~L.,    {Behroozi} P.~S.,
  2013, \apj, 771, 30

\bibitem[\protect\citeauthoryear{{Riebe}, {Partl}, {Enke}, {Forero-Romero},
  {Gottl{\"o}ber}, {Klypin}, {Lemson}, {Prada}, {Primack}, {Steinmetz} \&
  {Turchaninov}}{{Riebe} et~al.}{2013}]{riebe_etal13}
{Riebe} K.,  {Partl} A.~M.,  {Enke} H.,  {Forero-Romero} J.,  {Gottl{\"o}ber}
  S.,  {Klypin} A.,  {Lemson} G.,  {Prada} F.,  {Primack} J.~R.,  {Steinmetz}
  M.,    {Turchaninov} V.,  2013, Astronomische Nachrichten, 334, 691

\bibitem[\protect\citeauthoryear{{Rodr{\'{\i}}guez-Puebla}, {Drory} \&
  {Avila-Reese}}{{Rodr{\'{\i}}guez-Puebla} et~al.}{2012}]{rod_puebla12}
{Rodr{\'{\i}}guez-Puebla} A.,  {Drory} N.,    {Avila-Reese} V.,  2012, \apj,
  756, 2

\bibitem[\protect\citeauthoryear{{Salim} et~al.,}{{Salim}
  et~al.}{2007}]{salim07}
{Salim} S.,  et~al., 2007, \apjs, 173, 267

\bibitem[\protect\citeauthoryear{{Seljak}}{{Seljak}}{2000}]{seljak00}
{Seljak} U.,  2000, \mnras, 318, 203

\bibitem[\protect\citeauthoryear{{Shankar}, {Lapi}, {Salucci}, {De Zotti} \&
  {Danese}}{{Shankar} et~al.}{2006}]{shankar_etal06}
{Shankar} F.,  {Lapi} A.,  {Salucci} P.,  {De Zotti} G.,    {Danese} L.,  2006,
  \apj, 643, 14

\bibitem[\protect\citeauthoryear{{Sheth} \& {Tormen}}{{Sheth} \&
  {Tormen}}{2004}]{sheth_tormen04}
{Sheth} R.~K.,  {Tormen} G.,  2004, \mnras, 350, 1385

\bibitem[\protect\citeauthoryear{{Skibba}, {Sheth}, {Connolly} \&
  {Scranton}}{{Skibba} et~al.}{2006}]{skibba06}
{Skibba} R.,  {Sheth} R.~K.,  {Connolly} A.~J.,    {Scranton} R.,  2006,
  \mnras, 369, 68

\bibitem[\protect\citeauthoryear{{Skibba} et~al.,}{{Skibba}
  et~al.}{2013}]{skibbamarkedCF13}
{Skibba} R.~A.,  et~al., 2013, \mnras, 429, 458

\bibitem[\protect\citeauthoryear{{Skibba} \& {Sheth}}{{Skibba} \&
  {Sheth}}{2009a}]{skibba_sheth09}
{Skibba} R.~A.,  {Sheth} R.~K.,  2009a, \mnras, 392, 1080

\bibitem[\protect\citeauthoryear{{Skibba} \& {Sheth}}{{Skibba} \&
  {Sheth}}{2009b}]{skibba09a}
{Skibba} R.~A.,  {Sheth} R.~K.,  2009b, \mnras, 392, 1080

\bibitem[\protect\citeauthoryear{{Stein} \& {Soifer}}{{Stein} \&
  {Soifer}}{1983}]{stein_soifer83}
{Stein} W.~A.,  {Soifer} B.~T.,  1983, \araa, 21, 177

\bibitem[\protect\citeauthoryear{{Tal} et~al.,}{{Tal}
  et~al.}{2014}]{tal_etal14}
{Tal} T.,  et~al., 2014, ArXiv e-prints

\bibitem[\protect\citeauthoryear{{Tasitsiomi}, {Kravtsov}, {Wechsler} \&
  {Primack}}{{Tasitsiomi} et~al.}{2004}]{tasitsiomi_etal04}
{Tasitsiomi} A.,  {Kravtsov} A.~V.,  {Wechsler} R.~H.,    {Primack} J.~R.,
  2004, \apj, 614, 533

\bibitem[\protect\citeauthoryear{{Teyssier}, {Moore}, {Martizzi}, {Dubois} \&
  {Mayer}}{{Teyssier} et~al.}{2011}]{teyssier11}
{Teyssier} R.,  {Moore} B.,  {Martizzi} D.,  {Dubois} Y.,    {Mayer} L.,  2011,
  \mnras, 414, 195

\bibitem[\protect\citeauthoryear{{Tinker} et~al.,}{{Tinker}
  et~al.}{2013}]{tinker_etal13}
{Tinker} J.~L.,  et~al., 2013, \apj, 778, 93

\bibitem[\protect\citeauthoryear{{Trujillo-Gomez}, {Klypin}, {Primack} \&
  {Romanowsky}}{{Trujillo-Gomez} et~al.}{2011}]{trujillo_gomez11}
{Trujillo-Gomez} S.,  {Klypin} A.,  {Primack} J.,    {Romanowsky} A.~J.,  2011,
  \apj, 742, 16

\bibitem[\protect\citeauthoryear{{Vale} \& {Ostriker}}{{Vale} \&
  {Ostriker}}{2004}]{vale_ostriker04}
{Vale} A.,  {Ostriker} J.~P.,  2004, \mnras, 353, 189

\bibitem[\protect\citeauthoryear{{Vale} \& {Ostriker}}{{Vale} \&
  {Ostriker}}{2006}]{vale_ostriker06}
{Vale} A.,  {Ostriker} J.~P.,  2006, \mnras, 371, 1173

\bibitem[\protect\citeauthoryear{{van den Bosch} et~al.,}{{van den Bosch}
  et~al.}{2008}]{vdbosch_08}
{van den Bosch} F.~C.,  et~al., 2008, \mnras, 387, 79

\bibitem[\protect\citeauthoryear{{van den Bosch}, {More}, {Cacciato}, {Mo} \&
  {Yang}}{{van den Bosch} et~al.}{2013}]{vdBosch13}
{van den Bosch} F.~C.,  {More} S.,  {Cacciato} M.,  {Mo} H.,    {Yang} X.,
  2013, \mnras, 430, 725

\bibitem[\protect\citeauthoryear{{Wang}, {Farrah}, {Oliver}, {Amblard}, {Bock},
  {Conley}, {Cooray}, {Halpern}, {Heinis}, {Ibar}, {Ilbert}, {Ivison},
  {Marsden}, {Roseboom}, {Rowan-Robinson}, {Schulz}, {Smith}, {Viero} \&
  {Zemcov}}{{Wang} et~al.}{2012}]{wang_etal12}
{Wang} L.,  {Farrah} D.,  {Oliver} S.~J.,  {Amblard} A.,  {Bock} J.,  {Conley}
  A.,  {Cooray} A.,  {Halpern} M.,  {Heinis} S.,  {Ibar} E.,  {Ilbert} O.,
  {Ivison} R.~J.,  {Marsden} G.,  {Roseboom} I.~G.,  {Rowan-Robinson} M.,
  {Schulz} B.,  {Smith} A.~J.,  {Viero} M.,    {Zemcov} M.,  2012, ArXiv:
  1203.5828

\bibitem[\protect\citeauthoryear{{Wang} \& {Jing}}{{Wang} \&
  {Jing}}{2010}]{wang_jing10}
{Wang} L.,  {Jing} Y.~P.,  2010, \mnras, 402, 1796

\bibitem[\protect\citeauthoryear{{Watson}, {Berlind}, {McBride}, {Hogg} \&
  {Jiang}}{{Watson} et~al.}{2012}]{watson_etal12a}
{Watson} D.~F.,  {Berlind} A.~A.,  {McBride} C.~K.,  {Hogg} D.~W.,    {Jiang}
  T.,  2012, \apj, 749, 83

\bibitem[\protect\citeauthoryear{{Watson}, {Berlind}, {McBride} \&
  {Masjedi}}{{Watson} et~al.}{2010}]{watson_etal10}
{Watson} D.~F.,  {Berlind} A.~A.,  {McBride} C.~K.,    {Masjedi} M.,  2010,
  \apj, 709, 115

\bibitem[\protect\citeauthoryear{{Watson}, {Berlind} \& {Zentner}}{{Watson}
  et~al.}{2012}]{watson_etal12b}
{Watson} D.~F.,  {Berlind} A.~A.,    {Zentner} A.~R.,  2012, \apj, 754, 90

\bibitem[\protect\citeauthoryear{{Watson} \& {Conroy}}{{Watson} \&
  {Conroy}}{2013}]{watson_conroy13}
{Watson} D.~F.,  {Conroy} C.,  2013, \apj, 772, 139

\bibitem[\protect\citeauthoryear{{Wechsler}, {Bullock}, {Primack}, {Kravtsov}
  \& {Dekel}}{{Wechsler} et~al.}{2002}]{wechsler02}
{Wechsler} R.~H.,  {Bullock} J.~S.,  {Primack} J.~R.,  {Kravtsov} A.~V.,
  {Dekel} A.,  2002, \apj, 568, 52

\bibitem[\protect\citeauthoryear{{Wechsler}, {Zentner}, {Bullock}, {Kravtsov}
  \& {Allgood}}{{Wechsler} et~al.}{2006}]{wechsler06}
{Wechsler} R.~H.,  {Zentner} A.~R.,  {Bullock} J.~S.,  {Kravtsov} A.~V.,
  {Allgood} B.,  2006, \apj, 652, 71

\bibitem[\protect\citeauthoryear{{Weinmann}, {Kauffmann}, {van den Bosch},
  {Pasquali}, {McIntosh}, {Mo}, {Yang} \& {Guo}}{{Weinmann}
  et~al.}{2009}]{weinmann09}
{Weinmann} S.~M.,  {Kauffmann} G.,  {van den Bosch} F.~C.,  {Pasquali} A.,
  {McIntosh} D.~H.,  {Mo} H.,  {Yang} X.,    {Guo} Y.,  2009, \mnras, 394, 1213

\bibitem[\protect\citeauthoryear{{Weinmann}, {van den Bosch}, {Yang} \&
  {Mo}}{{Weinmann} et~al.}{2006}]{weinmann06b}
{Weinmann} S.~M.,  {van den Bosch} F.~C.,  {Yang} X.,    {Mo} H.~J.,  2006,
  \mnras, 366, 2

\bibitem[\protect\citeauthoryear{{Wetzel}, {Tinker} \& {Conroy}}{{Wetzel}
  et~al.}{2012}]{wetzel_etal11}
{Wetzel} A.~R.,  {Tinker} J.~L.,    {Conroy} C.,  2012, \mnras, 424, 232

\bibitem[\protect\citeauthoryear{{Wetzel}, {Tinker}, {Conroy} \& {van den
  Bosch}}{{Wetzel} et~al.}{2013}]{wetzel_etal12b}
{Wetzel} A.~R.,  {Tinker} J.~L.,  {Conroy} C.,    {van den Bosch} F.~C.,  2013,
  \mnras, 432, 336

\bibitem[\protect\citeauthoryear{{Wheeler}, {Phillips}, {Cooper},
  {Boylan-Kolchin} \& {Bullock}}{{Wheeler} et~al.}{2014}]{wheeler_etal14}
{Wheeler} C.,  {Phillips} J.~I.,  {Cooper} M.~C.,  {Boylan-Kolchin} M.,
  {Bullock} J.~S.,  2014, ArXiv:1402.1498

\bibitem[\protect\citeauthoryear{{Wyder} et~al.,}{{Wyder}
  et~al.}{2007}]{wyder07}
{Wyder} T.~K.,  et~al., 2007, \apjs, 173, 293

\bibitem[\protect\citeauthoryear{{Yang}, {Mo} \& {van den Bosch}}{{Yang}
  et~al.}{2003}]{yang03}
{Yang} X.,  {Mo} H.~J.,    {van den Bosch} F.~C.,  2003, \mnras, 339, 1057

\bibitem[\protect\citeauthoryear{{Yang}, {Mo}, {van den Bosch}, {Zhang} \&
  {Han}}{{Yang} et~al.}{2012}]{yang12}
{Yang} X.,  {Mo} H.~J.,  {van den Bosch} F.~C.,  {Zhang} Y.,    {Han} J.,
  2012, \apj, 752, 41

\bibitem[\protect\citeauthoryear{{York} et~al.,}{{York}
  et~al.}{2000}]{york00a}
{York} D.~G.,  et~al., 2000, \aj, 120, 1579

\bibitem[\protect\citeauthoryear{{Zehavi} et~al.,}{{Zehavi}
  et~al.}{2002}]{zehavi02}
{Zehavi} I.,  et~al., 2002, \apj, 571, 172

\bibitem[\protect\citeauthoryear{{Zehavi} et~al.,}{{Zehavi}
  et~al.}{2005}]{zehavi05a}
{Zehavi} I.,  et~al., 2005, \apj, 630, 1

\bibitem[\protect\citeauthoryear{{Zehavi} et~al.,}{{Zehavi}
  et~al.}{2011}]{zehavi11}
{Zehavi} I.,  et~al., 2011, \apj, 736, 59

\bibitem[\protect\citeauthoryear{{Zentner}, {Hearin} \& {van den
  Bosch}}{{Zentner} et~al.}{2014}]{zentner_etal14}
{Zentner} A.~R.,  {Hearin} A.~P.,    {van den Bosch} F.~C.,  2014, \mnras, 443,
  3044

\bibitem[\protect\citeauthoryear{{Zhang}, {Yang}, {Wang}, {Wang}, {Mo} \& {van
  den Bosch}}{{Zhang} et~al.}{2013}]{zhang_etal13}
{Zhang} Y.,  {Yang} X.,  {Wang} H.,  {Wang} L.,  {Mo} H.~J.,    {van den Bosch}
  F.~C.,  2013, \apj, 779, 160

\bibitem[\protect\citeauthoryear{{Zheng}, {Berlind}, {Weinberg}, {Benson},
  {Baugh}, {Cole}, {Dav{\'e}}, {Frenk}, {Katz} \& {Lacey}}{{Zheng}
  et~al.}{2005}]{zheng05}
{Zheng} Z.,  {Berlind} A.~A.,  {Weinberg} D.~H.,  {Benson} A.~J.,  {Baugh}
  C.~M.,  {Cole} S.,  {Dav{\'e}} R.,  {Frenk} C.~S.,  {Katz} N.,    {Lacey}
  C.~G.,  2005, \apj, 633, 791

\end{thebibliography}


\appendix


\section{SDSS Galaxy Clustering and Galaxy-Galaxy Lensing Measurements}
\label{app:tables}

\begin{table*}
  \caption{{\bf SDSS PROJECTED CORRELATION FUNCTION MEASUREMENTS:
      STAR-FORMING GALAXIES.}
    The first column is the mean radii of
    galaxies in each logarithmic bin in units of $\Mpc.$  Additional
    columns show the projected correlation function, $\wprp$, for star
    forming ($\mathrm{log_{10}(sSFR)} > -11$) galaxies for three
    stellar mass, volume-limited threshold samples:
    $\log_{10}(\Mstar)>[9.8,10.2,10.6]$.  Errors are computed from
    jackknife resampling of 50 equal-area regions on the sky, and the
    diagonal terms of the error covariance matrix are given in the
    parenthesis.}
\begin{center}
  \begin{tabular}{@{}ccccccc}
\\ \hline \hline
    $r_\mathrm{p}$ & $9.8$ & $10.2$ & $10.6$ \\ \hline
 0.121 & 167.63    (9.52) & 212.20   (15.74) &376.24   (68.50)  \\
 0.173 & 137.91    (7.47) & 169.69   (12.99) & 236.99   (30.19) \\
 0.247 & 109.89    (5.66) & 122.74   (7.31) & 171.51   (23.77)  \\
 0.352 &  92.71    (4.97) & 106.72    (7.02) & 114.35   (17.39) \\
 0.501 &  70.63    (3.61) &  79.28    (4.59) &  79.55   (10.41) \\
 0.714 &  61.20    (3.19) &  67.80    (3.71) &  71.64   (6.13)  \\
 1.017 &  46.95    (2.62) &  53.73    (3.19) &  54.41    (4.99) \\
 1.448 &  37.27    (2.40) &  43.11    (3.20) &  47.91    (4.73) \\
 2.060 &  29.37    (2.38) &  34.29    (3.03) &  36.58    (3.86) \\
 2.934 &  24.14    (2.29) &  27.45    (2.89) &  28.17    (3.15) \\
 4.178 &  19.97    (2.11) &  22.51   (2.60) &  23.54    (2.99)  \\
 5.946 &  15.25    (1.98) &  17.50    (2.57) & 17.63    (2.64)  \\
 8.467 &  10.67    (1.55) &  11.72    (1.97) &  12.34    (2.14) \\
12.056 &   7.41    (1.23) &   8.05    (1.44) &  8.21    (1.70) \\
17.163 &   4.16    (1.06) &   4.28    (1.20) &   4.34    (1.53) \\ \hline

  \end{tabular}
  \label{tab:wp_thresholds_SF}
\end{center}
\end{table*}


\begin{table*}
  \caption{{\bf SDSS PROJECTED CORRELATION FUNCTION MEASUREMENTS: QUENCHED GALAXIES.}
    Same as Table~\ref{tab:wp_thresholds_QUENCHED}, but for the quenched galaxy sample: $\mathrm{log_{10}(sSFR)} < -11$.}
\begin{center}
  \begin{tabular}{@{}ccccccc}
\\ \hline \hline
    $r_\mathrm{p}$ & $9.8$ & $10.2$ & $10.6$ \\ \hline
 0.121 & 937.41   (55.17)  & 952.28   (56.54) & 906.32   (50.58)\\
 0.173 & 714.72   (42.57) & 713.58   (44.09) & 603.32   (36.62)\\
 0.247 & 552.16   (40.38) & 551.71   (40.77) & 460.28   (28.24)\\
 0.352 & 414.42   (35.66) & 411.74   (33.96) & 346.20   (25.24)\\
 0.501 & 312.35   (28.37) & 306.81   (26.41) & 254.14   (20.73)\\
 0.714 & 226.84   (23.97) & 220.18   (22.02) & 186.01   (17.48)\\
 1.017 & 154.15   (18.02) & 150.92   (16.68) & 127.03   (13.17)\\
 1.448 & 103.35   (12.78) & 100.13   (11.64) &  88.78    (9.53)\\
 2.060 &  72.28   (10.07) &  70.81    (9.27) &  65.63    (7.43)\\
 2.934 &  52.52   (7.36) &  51.86    (6.84) &  51.03    (6.09)\\
 4.178 &  38.36    (5.69) &  37.86    (5.22) &  36.04    (4.18)\\
 5.946 &  27.48    (4.39) &  26.73    (4.17) &  25.88    (3.70)\\
 8.467 &  19.03    (3.18) &  18.83    (3.08) &  18.99    (2.80)\\
12.056 &  12.90    (2.17) &  12.78    (2.11) &  12.64    (2.08)\\
17.163 &   7.73    (1.66) &   7.53    (1.66) &   7.29    (1.66)\\ \hline

  \end{tabular}
  \label{tab:wp_thresholds_QUENCHED}
\end{center}
\end{table*}


\begin{table*}
  \label{tab:gg_thresholds_SF}
  \caption{{\bf SDSS GALAXY-GALAXY LENSING MEASUREMENTS: STAR-FORMING GALAXIES.}
    The first column is the mean radii of galaxies in each logarithmic bin, in units of $\kpc.$  Subsequent columns show the galaxy-galaxy lensing signal, $\Delta\Sigma$, in units of $\Msun\mathrm{pc}^{-2}$ for the same three stellar mass, volume-limited threshold samples. Errors in the parenthesis are derived from dividing the survey area into 200 bootstrap subregions and generating 500 bootstrap-resampled data sets.}
  \begin{tabular}{@{}ccccccc}
\\ \hline \hline
    R & $9.8$ & $10.2$ & $10.6$ \\ \hline
  31.87  & 25.20   (20.58) &   27.49   (26.79) &  -25.63   (46.62) \\
  38.93  & 24.35   (16.23) &   33.66   (21.47) &   76.67   (37.55) \\
  47.54  & 53.35   (12.37) &   58.77   (15.59) &   79.43   (31.38) \\
  58.07  & 25.61   (10.94) &   31.28   (14.38) &   28.00   (28.56) \\
  70.93  & 22.93    (9.57) &   26.64   (11.99) &   18.30   (21.53) \\
  86.63  & 11.06    (7.23) &    7.53    (9.68) &   28.00   (18.08) \\
 105.81  & 17.59    (5.64) &   11.42    (7.14) &   26.59   (11.84) \\
 129.24  &  1.95    (5.41) &   -1.28    (7.41) &   11.41   (11.80) \\
 157.85  &  3.16    (4.01) &   10.58    (5.29) &   21.32    (9.11) \\
 192.80  &  2.66    (3.00) &    2.05    (3.89) &    1.61    (7.33) \\
 235.49  &  3.35    (2.85) &    8.51    (3.69) &    7.15    (6.87) \\
 287.63  &  0.41    (2.42) &    0.33    (3.15) &    1.19    (5.76) \\
 351.31  &  2.26    (1.90) &    1.80    (2.42) &  -2.55    (4.18)  \\
 429.09  &  1.22    (1.71) &    2.31    (2.34) &    5.88    (3.87) \\
 524.09  &  1.69    (1.22) &    1.59    (1.58) &    1.33    (3.19) \\
 640.13  &  1.92    (1.04) &    0.90    (1.32) &   -0.21    (2.44) \\
 781.85  &  2.14    (0.83) &    1.52    (1.21) &   -1.26    (2.10) \\
 954.96  &  1.28    (0.68) &    1.65    (0.92) &    2.61    (1.50) \\
1166.39  &  1.27    (0.63) &    1.62    (0.80) &    0.54    (1.51) \\
1424.63  &  1.03    (0.53) &    1.05    (0.66) &    3.27    (1.22) \\
1740.04  &  1.15    (0.40) &    1.75    (0.53) &    2.26    (0.96) \\
2125.29  &  0.87    (0.33) &    0.96    (0.49) &    0.76    (0.84) \\
2595.84  &  0.87    (0.31) &    0.78    (0.40) &    0.51    (0.70) \\
3170.56  &  0.52    (0.23) &    0.65    (0.31) &    0.49    (0.55) \\ \hline
  \end{tabular}
\end{table*}

\begin{table*}
  \caption{{\bf SDSS GALAXY-GALAXY LENSING MEASUREMENTS: QUENCHED GALAXIES.}
    Same as Table~3, but for the quenched galaxy sample.}
  \begin{tabular}{@{}ccccccc}
\\ \hline \hline
    R & $9.8$ & $10.2$ & $10.6$ \\ \hline
  31.87  &   63.93   (20.39) &      67.10   (21.61) &    108.45   (29.36)\\
  38.93  &   19.61   (19.42) &      23.82   (20.14) &      38.28   (24.27)\\
  47.54  &   49.80   (14.42) &      58.03   (15.50) &     62.67   (19.85)\\
  58.07  &   35.86   (10.89) &      38.84   (11.68) &      34.55   (14.83)\\
  70.93  &  16.36    (9.28) &      20.42    (9.48) &     28.13   (13.60)\\
  86.63  &  19.23    (7.35) &       16.20    (7.49) &      20.33   (10.20)\\
 105.81  &   16.45    (6.09) &      14.62    (6.42) &      20.09    (8.52)\\
 129.24  &   19.80    (5.42) &      17.45    (5.33) &      23.94    (6.82)\\
 157.85  &   7.47    (4.47) &       6.47    (4.44) &       9.73    (5.47)\\
 192.80  &   10.42    (4.11) &      12.09    (4.23) &       14.98    (5.39)\\
 235.49  &   6.39    (3.07) &       5.33    (3.20) &       6.38    (3.99)\\
 287.63  &   11.60    (2.22) &      12.45    (2.38) &       17.15    (3.17)\\
 351.31  &    6.31    (1.83) &        6.21    (1.87) &      8.21    (2.61)\\
 429.09  &   6.84    (1.59) &        7.63    (1.58) &        8.59    (2.10)\\
 524.09  &    6.16    (1.31) &        6.34    (1.32) &        6.19    (1.82)\\
 640.13  &    7.25    (1.20) &       7.37    (1.25) &      7.32    (1.58)\\
 781.85  &    4.93    (1.01) &        4.64    (1.08) &       6.42    (1.37)\\
 954.96  &    5.61    (0.85) &        5.50    (0.87) &        4.10    (1.11)\\
1166.39  &    4.84    (0.81) &        5.15    (0.83) &        5.25    (0.96)\\
1424.63  &    3.38    (0.65) &        3.35    (0.66) &        3.29    (0.72)\\
1740.04  &    3.70    (0.66) &        3.69    (0.65) &        3.26    (0.73)\\
2125.29  &   2.65    (0.53) &       2.42    (0.52) &        2.21    (0.59)\\
2595.84  &    2.35    (0.49) &        2.31    (0.50) &        1.87    (0.52)\\
3170.56  &    2.26    (0.42) &       2.13    (0.42) &        1.71    (0.39)\\ \hline
  \end{tabular}
  \label{tab:gg_thresholds_QUENCHED}
\end{table*}


\end{document}